\pdfoutput=1
\documentclass[reprint, amsmath, amssymb, aps, pra, longbibliography, floatfix]{revtex4-2}

\usepackage{subfigure}
\usepackage{amsmath}
\usepackage{mathtools}
\usepackage{physics}
\usepackage{enumitem}
\usepackage{graphicx}
\usepackage{dcolumn}
\usepackage{bm}
\usepackage{placeins}
\usepackage[colorlinks=true,allcolors=blue]{hyperref}

\begin{document}

\title{Static entanglement structure and Bell-state preparation in the tripartite quantum Rabi model}
\thanks{Author's accepted manuscript. Published in \href{https://doi.org/10.1103/pd42-lb7j}{Phys. Rev. A \textbf{114}, 053722 (2026)}. \copyright~2026 American Physical Society.}

\author{Li-Li Gao}
\author{Zi-Xuan Chen}
\author{Long-Jie Li}
\author{Zi-Cheng Liu}
\author{A-Long Zhou}
\author{Chuan-Cun Shu}
\author{Zi-Min Li}
\email{zimin.li@csu.edu.cn}

\affiliation{Institute of Quantum Physics, Hunan Key Laboratory of Nanophotonics and Devices, Hunan Key Laboratory of Super-Microstructure and Ultrafast Process, School of Physics, Central South University, Changsha 410083, China}

\date{December 3, 2026}

\begin{abstract}
The tripartite quantum Rabi model couples two qubits to a bosonic mode through a collective spin-oscillator interaction, providing a simple setting for studying two-qubit entanglement.
In the zero-detuning limit, the triplet part of the spectrum splits into branches with zero and maximal entanglement, while the antisymmetric singlet ladder remains exactly decoupled.
Within the triplet sector, finite detuning turns the crossings between these branches into avoided crossings and redistributes this entanglement.
We identify an eigenbranch whose entanglement grows from nearly zero to a nearly maximal value through such avoided-crossing mixing.
The weak-coupling level ordering yields a simple analytic criterion for whether this eigenbranch has a separable weak-coupling endpoint.
A three-state effective model explains how the Bell-state component becomes dominant as the coupling increases.
We further compare linear and local adiabatic ramps of the collective coupling under common constraints, measuring the time needed to prepare a Bell-like state on the selected eigenbranch.
These results show how collective spin-oscillator coupling reorganizes spectral entanglement and connects static branch structure to finite-time Bell-state preparation.
\end{abstract}

\maketitle

\section{Introduction}

Quantum entanglement is a central feature of composite quantum systems and a key resource in quantum information science~\cite{Horodecki2009Entanglement}.
For two qubits, Bell states provide the simplest maximally entangled targets.
They are important for communication, gates, and state preparation and provide a simple reference for examining how correlations are organized in a physical system.
A basic question is therefore how a concrete model generates, rearranges, and controls two-qubit entanglement.

Light-matter interaction provides a versatile way to control entanglement between qubits~\cite{Raimond2001CavityEntanglement,Ficek2002TwoAtomEntanglement,Fink2009DressedCollective}.
In one common approach, the field mode is used as a mediator or an external control channel, for example in cavity-induced qubit-qubit interactions and adiabatic or shortcut-based state-preparation protocols.
A different approach is to treat the field and the qubits as a single quantum system and study the entanglement carried by its eigenstates.
Rabi-type models provide a basic setting for this viewpoint~\cite{FornDiaz2019USC,Kockum2019USC,Braak2011Integrability,BraakChenBatchelorSolano2016Rabi80,Xie2017RabiReview}.
In two-qubit and multiqubit Rabi models, symmetry decompositions, exact and quasi-exact states, exceptional states, dark-like states, and entanglement dynamics have been used to understand how entanglement is arranged in the spectrum~\cite{Agarwal2012,Chilingaryan2013TwoQubitRabi,Wang2014TwoQubit,Duan2015TwoQubit,Mao2015a,Mao2026BipartiteTripartite,Peng2014TwoQubitSolution,Peng2015AlgebraicStructure,Peng2017DarkLike,Sun2020Polaron,Grimaudo2023QPT,Grimaudo2024Thermodynamic}.
Bias or asymmetry terms can further modify this structure~\cite{Liu2017AsymPolaron,LiBatchelor2015Exceptional,LiFerri2021Nonorthogonal,LiBatchelor2021Hidden,LiBatchelor2021GAA,LiFerriTilbrook2021GAAAQRM,Shi2022EntanglementResonance}.
Adiabatic Bell-state generation has also been proposed using finite-photon dark states of multiqubit and anisotropic two-qubit Rabi models~\cite{Peng2021OnePhotonMMQRM,Xie2023UltrafastBell}.
Petiziol et al. studied adiabatic entanglement generation for two qubits dispersively coupled to a common resonator and accelerated the preparation using effective counterdiabatic driving~\cite{Petiziol2019}.
Their protocol sweeps the qubit frequencies toward mutual resonance and implements the acceleration through rapid modulation of the individual qubit--resonator couplings.

In many two-qubit and multiqubit Rabi-type models, the interaction couples the field mode to each qubit through a single-qubit spin operator.
The tripartite quantum Rabi model (TQRM) introduces a different interaction: the field mode couples directly to a joint two-qubit operator~\cite{Hamlyn2024}.
The two-qubit state then enters the light-matter interaction as a collective degree of freedom.
This structure makes collective spin states the natural basis for organizing the spectrum.
It also raises the question of whether the eigenstates carry two-qubit entanglement in a form specific to the tripartite coupling.
In this work, we study this spectral entanglement structure and identify the features produced by the collective interaction.

In the static spectrum, the zero-detuning symmetric limit provides a useful reference point.
In this limit, the triplet part of the spectrum separates into branches with zero and maximal reduced two-qubit entanglement, while the antisymmetric singlet ladder remains exactly decoupled.
Finite detuning mixes these triplet branches.
As a spectral consequence, the level crossings between these triplet branches in the symmetric limit are lifted into avoided crossings.
Owing to the mixing, the eigenstate entanglement is no longer restricted to the two fixed values of the symmetric limit.
The change is most pronounced near the avoided crossings, where the branch mixing is strongest.
Specifically, for suitable parameters, some low-energy eigenbranches start with weak entanglement at small coupling and become nearly maximally entangled as the coupling strength is increased.
The weak-coupling level ordering gives a simple analytic criterion for whether the selected eigenbranch has a separable weak-coupling endpoint.
The avoided-crossing mixing establishes the exact branch connection, while a three-state displaced-frame model captures the associated spin-weight transfer.

The static route suggests a direct preparation protocol.
Starting from the weak-coupling end of the selected branch, the coupling strength is increased in time to follow the branch through the avoided-crossing structure.
When this evolution is sufficiently adiabatic, the reduced two-qubit state approaches a Bell-state target at the strong-coupling end.
We compare linear and local adiabatic schedules under common bounds on coupling strength and available time.
Both protocols keep the qubit parameters fixed and vary only the collective coupling.
Preparation is assessed at the first time that the evolved state's Bell fidelity, its fidelity with the selected eigenbranch, and the target eigenstate's own Bell fidelity all meet the prescribed threshold.
The resulting maps link the analytic entrance criterion to the coupling needed by the target eigenstate and the time needed to prepare it.
Representative trajectory tests and a quantitative comparison with related protocols establish the control consequences of this branch selection.

The paper is organized as follows.
Section~\ref{sec:model} defines the TQRM Hamiltonian, the two-qubit basis, and the triplet-sector structure used below.
Section~\ref{sec:static} analyzes the static entanglement structure, develops the three-state effective description, and derives the entrance criterion for the low-energy route.
Section~\ref{sec:bell} compares linear and local adiabatic ramps for Bell-state preparation and determines the coupling and time requirements.
Section~\ref{sec:discussion} discusses the experimental relevance, the range of applicability, and possible generalizations.
Section~\ref{sec:conclusion} concludes the paper.

\section{Model, Symmetry, and Effective Triplet Structure}
\label{sec:model}

We first define the rotated-basis TQRM Hamiltonian used throughout this work.
We then introduce a Bell basis for the two-qubit states, in which the singlet component separates from the coupled triplet sector.
Finally, we derive an effective three-state triplet model that captures the structure of the relevant low-energy spectrum.

\subsection{Tripartite quantum Rabi model}

We consider the Hamiltonian
\begin{equation}
H=\omega a^\dagger a-\Omega(\sigma_1^z+\sigma_2^z)+g(a+a^\dagger)\sigma_1^x\sigma_2^x+\epsilon(\sigma_1^x+\sigma_2^x),
\label{eq:Hamiltonian}
\end{equation}
where $\omega$ is the oscillator frequency, $2\Omega$ is the zero-bias qubit splitting, $g$ is the tripartite coupling strength, and $\epsilon$ is the detuning parameter of the original TQRM implementation.
The spin states $\ket{\uparrow}$ and $\ket{\downarrow}$ are eigenstates of the single-qubit $\sigma_i^z$ operators in Eq.~\eqref{eq:Hamiltonian}, with eigenvalues $+1$ and $-1$.
With this spin convention, the tunable detuning appears as the transverse single-spin term $\epsilon(\sigma_1^x+\sigma_2^x)$.
We use $g/\omega$ as the main control variable, while $\Omega/\omega$ and $\epsilon/\omega$ set the spectral organization and two-qubit entanglement structure.
We set $\hbar=1$ and measure energies in units of $\omega$.
For displaced oscillator states, we write $\alpha=g/\omega$ and $D(\beta)=\exp[\beta(a^\dagger-a)]$ for real $\beta$.

\begin{figure}[tbp]
\centering
\includegraphics[width=\linewidth]{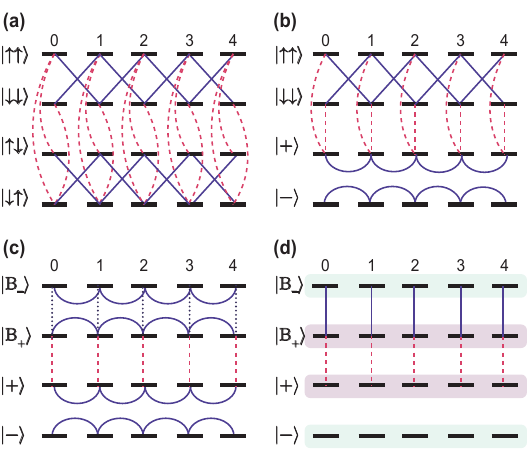}
\caption{
Schematic organization of the TQRM coupling structure.
Black horizontal segments denote oscillator levels.
In (a)--(c), the column labels $n$ denote bare Fock-state indices.
(a) Two-qubit $\sigma^z$ product-basis graph; purple solid links show the collective spin-oscillator term $g(a+a^\dagger)\sigma_1^x\sigma_2^x$, and red dashed links show the detuning-derived single-spin term $\epsilon(\sigma_1^x+\sigma_2^x)$.
(b) Reorganization of the one-excitation rows into the symmetric and antisymmetric states $\ket{+}$ and $\ket{-}$.
(c) Full Bell-basis graph, where the singlet row $\ket{-}$ is isolated and $\ket{+}$, $\ket{B_+}$, and $\ket{B_-}$ form the triplet network.
The purple curved links in (c) come from the row-preserving collective spin-oscillator term, the red dashed vertical links come from the detuning-derived term, and the purple dotted vertical links between $\ket{B_+}$ and $\ket{B_-}$ come from the qubit-splitting term $-\Omega(\sigma_1^z+\sigma_2^z)$.
(d) Fixed-$n$ effective triplet ladder after the displaced-frame reduction.
The column label $n$ denotes the transformed-frame Fock index retained in the diagonal Franck-Condon approximation.
The pale bands mark the two displacement classes: $D(-\alpha)\ket{n}$ for $\ket{+}$ and $\ket{B_+}$, and $D(+\alpha)\ket{n}$ for $\ket{B_-}$ and $\ket{-}$.
In (d), the red dashed link has strength $2\epsilon$, the solid purple link has strength $2\Omega G_n(\alpha)$, and the singlet row remains decoupled.
}
\label{fig:model_overview}
\end{figure}

The distinctive part of the model is the collective interaction
\begin{equation}
g(a+a^\dagger)\sigma_1^x\sigma_2^x,
\label{eq:collective_term}
\end{equation}
which couples the oscillator to a simultaneous flip of the two qubits.
We first view this coupling in the two-qubit $\sigma^z$ product basis $\{\ket{\uparrow\uparrow},\ket{\downarrow\downarrow},\ket{\uparrow\downarrow},\ket{\downarrow\uparrow}\}$.
In this basis, Fig.~\ref{fig:model_overview}(a) shows the tripartite term as purple solid links between rows with both spins flipped, while the factor $a+a^\dagger$ changes the oscillator index by one.
The detuning-derived term $\epsilon(\sigma_1^x+\sigma_2^x)$ appears as red dashed links, which flip one qubit at a time without changing the oscillator index.
Thus panel (a) shows the two elementary processes generated by the two non-diagonal parts of the Hamiltonian.
Panel (b) reorganizes the one-excitation rows into the symmetric and antisymmetric states
\begin{equation}
\ket{+}=\frac{\ket{\uparrow\downarrow}+\ket{\downarrow\uparrow}}{\sqrt2},
\qquad
\ket{-}=\frac{\ket{\uparrow\downarrow}-\ket{\downarrow\uparrow}}{\sqrt2}.
\label{eq:Bell_pm}
\end{equation}
In this basis, the antisymmetric row is already isolated, while the remaining rows form the coupled TQRM block.

\subsection{Bell basis and symmetry decomposition}

To make the two-qubit entanglement structure explicit, we complete the Bell basis by combining the parallel-spin states into
\begin{equation}
\ket{B_+}=\frac{\ket{\uparrow\uparrow}+\ket{\downarrow\downarrow}}{\sqrt2},
\qquad
\ket{B_-}=\frac{\ket{\uparrow\uparrow}-\ket{\downarrow\downarrow}}{\sqrt2}.
\label{eq:Bell_Bpm}
\end{equation}
Together with $\ket{+}$ and $\ket{-}$, these states split the spin space into a one-dimensional singlet sector $\mathcal S$ and a triplet sector $\mathcal T$,
\begin{equation}
\mathcal S=\mathrm{span}\{\ket{-}\},
\qquad
\mathcal T=\mathrm{span}\{\ket{+},\ket{B_+},\ket{B_-}\}.
\label{eq:singlet_triplet_spaces}
\end{equation}
For the oscillator part, we denote the bare Fock state by $\ket{n}$.
The decoupled singlet sector $\mathcal S$ then gives an independent displaced-oscillator ladder, with eigenstates $\ket{-}\otimes D(\alpha)\ket{n}$ and corresponding energies $n\omega-g^2/\omega$.
The remaining triplet sector $\mathcal T$ contains the coupled components $\ket{+}$, $\ket{B_+}$, and $\ket{B_-}$, and its explicit Bell-basis block is given in Appendix~\ref{app:bell_basis}.

In the full Bell basis, the collective flip operator $\sigma_1^x\sigma_2^x$ in the tripartite coupling is diagonal, with eigenvalue $+1$ for $\ket{+}$ and $\ket{B_+}$, and eigenvalue $-1$ for $\ket{-}$ and $\ket{B_-}$.
Panel (c) of Fig.~\ref{fig:model_overview} represents this diagonal action as a coupling graph.
In this graph, the collective spin-oscillator term preserves the Bell-state row and connects neighboring oscillator levels within that row, producing the purple curved links.
The detuning-derived term gives the red dashed links between $\ket{+}$ and $\ket{B_+}$, while the qubit-splitting term $-\Omega(\sigma_1^z+\sigma_2^z)$ gives the purple dotted vertical links between $\ket{B_+}$ and $\ket{B_-}$.
Thus panel (c) gives the exact Bell-basis coupling network, with the isolated singlet row separated from the three-row triplet block.

\subsection{Triplet chain and effective model}

The triplet network in Fig.~\ref{fig:model_overview}(c) is an exact Bell-basis representation of the Hamiltonian, with all bare oscillator levels $\ket{n}$ retained.
The same block can be rewritten in a displaced-oscillator frame, as detailed in Appendix~\ref{app:polaron} and represented schematically in panel (d).
In this frame, the bare-Fock hopping generated by $g(a+a^\dagger)$ is absorbed into the displacement.
For a Bell component with $\sigma_1^x\sigma_2^x$ eigenvalue $\eta=\pm1$, the tripartite term acts on the oscillator as the linear drive $\eta g(a+a^\dagger)$, which is absorbed by the displacement $D(-\eta\alpha)$ with $\alpha=g/\omega$.
In this frame, $\ket{+}$ and $\ket{B_+}$ carry $D(-\alpha)\ket{m}$, whereas $\ket{B_-}$ and the decoupled singlet row $\ket{-}$ carry $D(+\alpha)\ket{m}$.
The displacement classes differ structurally from those of an additive two-qubit Rabi coupling.
For the product operator, $(\sigma_1^x\sigma_2^x)^2=\mathbb I$, and completing the oscillator square therefore produces the common shift $-g^2/\omega$ for all spin components.
For the normalized additive operator $(\sigma_1^x+\sigma_2^x)/2$, the square is $(\mathbb I+\sigma_1^x\sigma_2^x)/2$, and the corresponding shift depends on the spin state.
In particular, the product coupling displaces the singlet oscillator, whereas the additive coupling annihilates the singlet.
These are features of the TQRM interaction introduced in Ref.~\cite{Hamlyn2024}; here they organize the Bell-basis branches used to derive the preparation route.
Within the triplet block, the red dashed vertical links connect $\ket{+}$ and $\ket{B_+}$ inside the same displacement class and have strength $2\epsilon$.
The solid vertical links connect $\ket{B_+}$ and $\ket{B_-}$ across the two displacement classes and therefore carry Franck-Condon matrix elements $\bra{m}D(2\alpha)\ket{m'}$.
For the fixed-$n$ reference, we project the Franck-Condon coupling onto the same oscillator index, $m=m'=n$.
This gives the diagonal Franck-Condon factor
\begin{equation}
G_n(\alpha)=e^{-2\alpha^2}L_n(4\alpha^2),
\qquad
\alpha=\frac{g}{\omega}.
\label{eq:Gn}
\end{equation}
Here $L_n$ is the Laguerre polynomial.
This diagonal-overlap reduction gives the fixed-$n$ ladder shown in Fig.~\ref{fig:model_overview}(d): the red dashed link has strength $2\epsilon$, the solid purple link has strength $2\Omega G_n(\alpha)$, and the singlet row remains disconnected.
The off-diagonal overlaps with $m\ne m'$, which connect different fixed-$n$ reference groups, remain part of the exact displaced-frame formulation given in Appendix~\ref{app:polaron}.

Within the same-$n$ ladder, the three triplet components form the chain
\begin{equation}
\ket{+} \xleftrightarrow{\,2\epsilon\,} \ket{B_+} \xleftrightarrow{\,2\Omega G_n(\alpha)\,} \ket{B_-},
\label{eq:triplet_chain}
\end{equation}
Thus $\ket{+}$ and $\ket{B_-}$ are mixed only through the intermediate component $\ket{B_+}$.

In the ordered basis
\begin{equation}
\{\ket{B_+},\ \ket{+},\ \ket{B_-}\}
\end{equation}
this chain is represented by the effective Hamiltonian
\begin{equation}
H_n^{\rm eff}
=
\left(n\omega-\frac{g^2}{\omega}\right)\mathbb I
+
2
\begin{pmatrix}
0 & \epsilon & -\Omega G_n(\alpha) \\
\epsilon & 0 & 0 \\
-\Omega G_n(\alpha) & 0 & 0
\end{pmatrix}.
\label{eq:Heff_n}
\end{equation}
The polaron transformation is exact, whereas Eq.~\eqref{eq:Heff_n} is an approximation: it discards couplings between different transformed-frame oscillator indices.
It describes the spin composition within a fixed-$n$ reference group; intergroup avoided crossings require the full Hamiltonian.
Its three eigenvalues are
\begin{equation}
\begin{aligned}
E_{n,0}^{\rm eff}&=n\omega-g^2/\omega,\\
E_{n,\pm}^{\rm eff}&=n\omega-g^2/\omega\pm2\Lambda_n(g),\\
\Lambda_n(g)&=\sqrt{\epsilon^2+\Omega^2G_n(\alpha)^2}.
\end{aligned}
\label{eq:effective_energies}
\end{equation}
Appendix~\ref{app:polaron} gives the complete normalized eigenbasis and quantitative comparisons with the full triplet Hamiltonian.

The effective Hamiltonian has one eigenstate with no $\ket{B_+}$ component,
\begin{equation}
\ket{D_n(g)}=
\frac{\Omega G_n(\alpha)\,\ket{+}+\epsilon\,\ket{B_-}}
{\sqrt{\epsilon^2+\Omega^2G_n(\alpha)^2}}.
\label{eq:Dn}
\end{equation}
The $\ket{B_+}$ row of Eq.~\eqref{eq:Heff_n} gives $2[\epsilon\,\Omega G_n(\alpha)-\Omega G_n(\alpha)\epsilon]=0$; this state therefore has eigenenergy $n\omega-g^2/\omega$.
The absence of the intermediate component means that $\ket{D_n(g)}$ tracks the redistribution of weight between $\ket{+}$ and $\ket{B_-}$ inside the effective chain.
To compare this fixed-$n$ reference with exact spin-oscillator eigenstates, we construct the corresponding full spin-oscillator state by assigning each component its displacement class,
\begin{equation}
\begin{aligned}
\ket{\mathcal D_n^{\rm full}(g)}
=&
\frac{1}{\sqrt{\epsilon^2+\Omega^2G_n(\alpha)^2}}
\Big[
\Omega G_n(\alpha)\ket{+}\otimes D(-\alpha)\ket{n}
\\
&\hspace{2.5em}
\quad+
\epsilon\ket{B_-}\otimes D(+\alpha)\ket{n}
\Big].
\end{aligned}
\label{eq:Dn_full}
\end{equation}
The reference fidelity is then defined as
\begin{equation}
F_{D_n}(g)=
\left|\left\langle \mathcal D_n^{\rm full}(g)\middle|\psi(g)\right\rangle\right|^2,
\label{eq:Dn_fidelity_full}
\end{equation}
where $\ket{\psi(g)}$ is the exact eigenstate being followed.

For $n=0$, the Franck-Condon factor
\begin{equation}
G_0(\alpha)=e^{-2\alpha^2}
\label{eq:G0}
\end{equation}
decreases monotonically with $g$.
For nonzero $\epsilon$, the $\ket{+}$ weight in $\ket{D_0(g)}$ is suppressed as the coupling grows, while the $\ket{B_-}$ weight becomes dominant.
This $n=0$ reference channel is the one relevant to the low-energy branch structure.

Appendix~\ref{app:polaron} tests the $n=0$ reference state against the second-lowest exact triplet eigenstate, whose connection to a separable entrance is established in Sec.~\ref{sec:static}.
On the 466 parameter pairs specified there, this eigenstate has $F_{D_0}\geq0.9517$ at $g/\omega=2$ and $F_{D_0}\geq0.9913$ at $g/\omega=2.5$.
These comparisons establish the usefulness of the large-coupling state reference on that sample.
The other states and the nearest spectral gap can have appreciably larger errors; all branch tracking and control calculations below therefore use the full Hamiltonian.

\section{Static Entanglement Structure and Selected Route}
\label{sec:static}

We now examine how the static eigenstates organize two-qubit entanglement as the coupling strength is varied.
The exactly decoupled singlet ladder forms an independent Bell-entangled family, while the coupled triplet branches contain the nontrivial entanglement reorganization.
To follow this reorganization, we use two solvable limits as endpoint labels for the exact branches.
At $g=0$, the oscillator and spin parts separate, while for large $g/\omega$ the spectrum is organized by displaced oscillator states.
The exact eigenbranches connecting these endpoint labels show where the two-qubit entanglement changes across the spectrum.

To quantify this entanglement, we trace out the oscillator from each exact spin-oscillator eigenstate and obtain the reduced two-qubit density matrix $\rho_{12}$.
We then use the concurrence of $\rho_{12}$ as the entanglement measure.
For a general $\rho_{12}$, the concurrence is
\begin{equation}
C(\rho_{12})=\max\{0,\lambda_1-\lambda_2-\lambda_3-\lambda_4\},
\label{eq:concurrence}
\end{equation}
where $\lambda_i$ are the square roots of the eigenvalues of
\begin{equation}
\rho_{12}(\sigma_1^y\otimes\sigma_2^y)\rho_{12}^{*}(\sigma_1^y\otimes\sigma_2^y)
\end{equation}
in descending order~\cite{Hill1997,Wootters1998EOF}.
For the states considered below, $C=0$ indicates a separable two-qubit state, while values close to unity indicate Bell-like entanglement.
Following $C$ along an exact branch then shows how the two-qubit entanglement changes across the spectrum.

\subsection{Two solvable endpoints}

At $g=0$, the tripartite coupling vanishes and the full Hamiltonian reduces to a sum of a bare harmonic oscillator and two identical one-spin Hamiltonians,
\begin{equation}
H(g=0)=\omega a^\dagger a+\sum_{j=1}^2
\left(-\Omega\sigma_j^z+\epsilon\sigma_j^x\right).
\end{equation}
The one-spin Hamiltonian is diagonalized by a rotation and has eigenvalues
$\pm\sqrt{\Omega^2+\epsilon^2}$.
Therefore, the exact eigenstates at this endpoint are product states of oscillator Fock states and spin eigenstates, with energies
\begin{equation}
E^{\rm L}_{n,0}=n\omega,
\qquad
E^{\rm L}_{n,\pm}=n\omega\pm2\sqrt{\Omega^2+\epsilon^2}.
\label{eq:g0_energies}
\end{equation}
The superscript ${\rm L}$ denotes the left, weak-coupling endpoint labels of the exact branches and fixes their ordering at $g=0$.

The other endpoint is the large-coupling limit, $g/\omega\gg1$, where the tripartite term $g(a+a^\dagger)\sigma_1^x\sigma_2^x$ controls the oscillator coordinate.
Writing $\alpha=g/\omega$, and fixing the eigenvalue $\eta=\pm1$ of $\sigma_1^x\sigma_2^x$, one obtains
\begin{equation}
\omega a^\dagger a+\eta g(a+a^\dagger)
=
\omega\left(a^\dagger+\eta\alpha\right)
\left(a+\eta\alpha\right)
-\frac{g^2}{\omega}.
\end{equation}
The oscillator states are therefore displaced Fock states $D(-\eta\alpha)\ket{n}$.
The two displaced oscillator ladders associated with $\eta=\pm1$ are separated by $2\alpha$, and the diagonal Franck-Condon overlap $G_n(\alpha)=\bra{n}D(2\alpha)\ket{n}$ vanishes as $g/\omega$ grows.
In the fixed-$n$ triplet model, the $\ket{B_+}\leftrightarrow\ket{B_-}$ coupling $2\Omega G_n(\alpha)$ is then suppressed, and the three levels with the same $n$ approach
\begin{equation}
\begin{aligned}
E^{\rm R}_{n,0} &\to n\omega-\frac{g^2}{\omega},\\
E^{\rm R}_{n,\pm} &\to n\omega-\frac{g^2}{\omega}\pm2|\epsilon|.
\end{aligned}
\label{eq:large_g_energies}
\end{equation}

The superscript ${\rm R}$ denotes the right, large-coupling endpoint labels.
The large-coupling ordering $E^{\rm R}_{n,-}<E^{\rm R}_{n,0}<E^{\rm R}_{n,+}$ is fixed within each $n$ group for finite $|\epsilon|$, independent of $\Omega$.
By contrast, the weak-$g$ ordering in Eq.~\eqref{eq:g0_energies} depends on $\Omega$ and $\epsilon$, especially when branches from neighboring oscillator indices are compared.
Together, the $g=0$ and large-coupling limits provide the two endpoint labels; the finite-$g$ spectrum determines how these labels are connected.

\begin{figure*}[tbp]
\centering
\includegraphics[width=\textwidth]{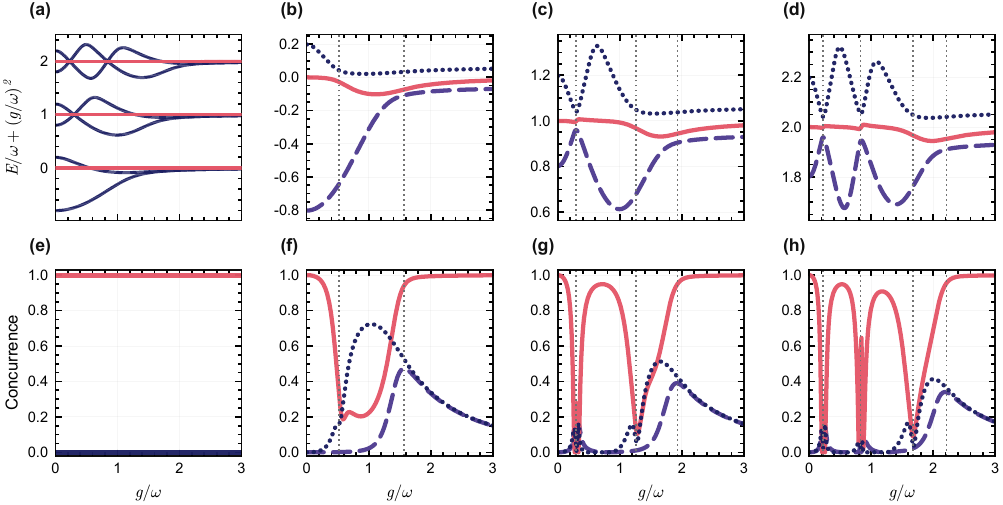}
\caption{
Static overview of the low-energy triplet structure at $\Omega/\omega=0.40$ and $\epsilon/\omega=0.03$.
The left column gives the zero-detuning reference.
In this column, purple curves denote QRM reference branches with $C=0$, while red horizontal lines denote DO reference branches with $C=1$.
The other columns show the finite-$\epsilon$ spectrum in the three groups labeled $n=0,1,2$.
The curves are exact tracked branches grouped by their large-$g$ displaced-oscillator limits, whose oscillator indices give the labels $n=0,1,2$.
For each group, the upper row shows the flattened energies and the lower row shows the concurrence of the same tracked branches.
In the finite-$\epsilon$ columns, purple dashed, red solid, and dark-purple dotted curves denote the $-$, $0$, and $+$ tracked branches, respectively.
Finite $\epsilon$ changes the concurrence mainly near avoided crossings.
The vertical dotted lines are guides to the avoided-crossing windows.
}
\label{fig:triplet_groups}
\end{figure*}

\subsection{\texorpdfstring{$\epsilon=0$}{epsilon=0} symmetric reference}

At the zero-detuning symmetric point $\epsilon=0$, the term $\epsilon(\sigma_1^x+\sigma_2^x)$ vanishes and the spin Hilbert space separates into two invariant subspaces.
One is the parallel-spin subspace
\begin{equation}
\mathcal H_{\parallel}=\mathrm{span}\{\ket{\uparrow\uparrow},\ket{\downarrow\downarrow}\}.
\end{equation}
The other is the one-excitation subspace
\begin{equation}
\mathcal H_{\mathrm{1ex}}=\mathrm{span}\{\ket{\uparrow\downarrow},\ket{\downarrow\uparrow}\}.
\end{equation}
These two subspaces give two reference entanglement classes in the symmetric spectrum.

In $\mathcal H_{\parallel}$, the qubit-splitting term distinguishes $\ket{\uparrow\uparrow}$ from $\ket{\downarrow\downarrow}$, while the tripartite coupling flips the two spins together.
With the pseudo-spin operators
\begin{equation}
\begin{aligned}
S^z&=\frac12\left(\ket{\uparrow\uparrow}\bra{\uparrow\uparrow}-\ket{\downarrow\downarrow}\bra{\downarrow\downarrow}\right),\\
S^x&=\frac12\left(\ket{\uparrow\uparrow}\bra{\downarrow\downarrow}+\ket{\downarrow\downarrow}\bra{\uparrow\uparrow}\right),
\end{aligned}
\end{equation}
the projected Hamiltonian becomes
\begin{equation}
H_{\mathrm{QRM}}=\omega a^\dagger a-4\Omega S^z+2g\,S^x(a^\dagger+a),
\label{eq:HQ}
\end{equation}
which is a quantum Rabi Hamiltonian for the pseudo-spin formed by the two parallel-spin states.
We refer to this invariant part of the spectrum as the QRM sector.

The vanishing concurrence in this sector follows from the parity symmetry of $H_{\mathrm{QRM}}$.
In a fixed QRM parity sector, an eigenstate can be written as
\begin{equation}
\ket{\psi_{\mathrm{QRM}}}
=
\ket{\uparrow\uparrow}\ket{\phi_{\mathrm e}}
+
\ket{\downarrow\downarrow}\ket{\phi_{\mathrm o}},
\end{equation}
or with the even and odd oscillator states interchanged.
Here $\ket{\phi_{\mathrm e}}$ and $\ket{\phi_{\mathrm o}}$ contain only even and odd Fock states, respectively, and hence satisfy $\langle\phi_{\mathrm e}|\phi_{\mathrm o}\rangle=0$.
When the oscillator is traced out, the off-diagonal element between $\ket{\uparrow\uparrow}$ and $\ket{\downarrow\downarrow}$ is proportional to this overlap and therefore vanishes.
The reduced two-qubit state is consequently an incoherent mixture of the two parallel-spin product states, which is separable and has
\begin{equation}
C_{\mathrm{QRM}}=0.
\end{equation}
Thus the QRM sector supplies the zero-concurrence reference at the symmetric point.

In $\mathcal H_{\mathrm{1ex}}$, the qubit-splitting term vanishes on each basis state,
\begin{equation}
(\sigma_1^z+\sigma_2^z)\ket{\uparrow\downarrow}=0,
\qquad
(\sigma_1^z+\sigma_2^z)\ket{\downarrow\uparrow}=0.
\end{equation}
The remaining spin dependence comes from $\sigma_1^x\sigma_2^x$, which exchanges the two one-excitation states and is diagonal in the Bell basis,
\begin{equation}
\sigma_1^x\sigma_2^x\ket{+}=\ket{+},
\qquad
\sigma_1^x\sigma_2^x\ket{-}=-\ket{-}.
\end{equation}
The eigenstates in this sector are therefore displaced-oscillator products,
\begin{equation}
\ket{+}\otimes D(-\alpha)\ket{n},
\qquad
\ket{-}\otimes D(+\alpha)\ket{n},
\label{eq:DOstates}
\end{equation}
with displaced-oscillator energy $n\omega-g^2/\omega$.
We refer to these displaced-oscillator product ladders as the displaced-oscillator (DO) sector.
Since each DO eigenstate is a product of a Bell spin state and an oscillator state, tracing out the oscillator leaves a pure Bell state and gives
\begin{equation}
C_{\mathrm{DO}}=1.
\end{equation}
The antisymmetric state $\ket{-}$ belongs to the exactly decoupled singlet ladder and remains Bell-entangled throughout.
By contrast, the symmetric state $\ket{+}$ remains in the triplet block, where finite $\epsilon$ can mix it with the QRM sector and produce nontrivial entanglement reorganization.

The triplet part of this QRM--DO classification is visible in the left column of Fig.~\ref{fig:triplet_groups}.
In Fig.~\ref{fig:triplet_groups}(a), the purple curves are QRM branches, while the red horizontal lines are the $\ket{+}$ DO branches.
The two sets of branches can cross because the QRM and DO sectors are decoupled at $\epsilon=0$; crossings within the QRM sector are protected by QRM parity.
The concurrence panel in Fig.~\ref{fig:triplet_groups}(e) shows the corresponding fixed entanglement structure: the purple QRM branches stay at $C=0$, whereas the red DO branches stay at $C=1$.
Thus, at $\epsilon=0$, level crossings occur without redistributing concurrence between the reference branches.
At the exact crossing points, the eigenbasis inside the degenerate subspace is not unique; the $C=0$ and $C=1$ assignments refer to the parity-adapted QRM states and the displaced Bell states used as the reference basis.

\subsection{Finite-\texorpdfstring{$\epsilon$}{epsilon} static overview}

At finite $\epsilon$, the crossings visible in Fig.~\ref{fig:triplet_groups}(a) are lifted into avoided-crossing windows.
Within the triplet block, the term $\epsilon(\sigma_1^x+\sigma_2^x)$ couples the symmetric one-excitation state $\ket{+}$ to the parallel-spin component $\ket{B_+}$.
The same term also breaks the QRM parity that protected crossings between QRM branches at $\epsilon=0$.
The finite-$\epsilon$ triplet spectrum therefore contains two classes of avoided crossings.
One lies between QRM and DO reference branches, and the other lies within the QRM branch family.

Figure~\ref{fig:triplet_groups} summarizes the finite-$\epsilon$ reshaping against the $\epsilon=0$ reference.
Every finite-$\epsilon$ energy curve is obtained by numerical diagonalization of the full triplet Hamiltonian and continuation of its eigenstates in $g$.
The concurrence is calculated after tracing the oscillator out of those same eigenstates.
Equation~\eqref{eq:Heff_n} supplies reference labels and an interpretation of their composition.
The left column gives the zero-detuning reference spectrum, where the low-energy branches separate visually from bottom to top into the $n=0,1,2$ groups at the plotted parameters.
The three finite-$\epsilon$ column pairs, Figs.~\ref{fig:triplet_groups}(b,f), \ref{fig:triplet_groups}(c,g), and \ref{fig:triplet_groups}(d,h), then show the corresponding $n=0,1,2$ groups.
The upper row plots the flattened energies $E/\omega+(g/\omega)^2$, which removes the common displaced-oscillator shift, while the lower row plots the concurrence of the corresponding tracked branches.
Within each finite-$\epsilon$ group, the purple dashed, red solid, and dark-purple dotted lines denote the same three branches in the energy and concurrence panels.
The vertical dotted guides mark the avoided-crossing windows, where level repulsion in the upper panels coincides with concurrence redistribution in the lower panels.
More precisely, the label $n$ is assigned by tracking each group to the large-$g$ side, where it connects to the corresponding fixed-$n$ displaced-oscillator structure.

The $n=0$ group gives the cleanest example of how a crossing in the $\epsilon=0$ reference is lifted at finite $\epsilon$.
In Fig.~\ref{fig:triplet_groups}(a), the corresponding QRM and DO branches cross because the two sectors are decoupled.
At finite $\epsilon$, the $n=0$ group is shown in Fig.~\ref{fig:triplet_groups}(b), where this crossing becomes an avoided crossing and the red middle branch is repelled from its neighboring branches inside the dotted window.
Figure~\ref{fig:triplet_groups}(f) shows the concurrence of the same three tracked branches.
The branch with high concurrence before the window loses much of that concurrence in the mixing region and recovers it after the avoided crossing.
The neighboring branches change in the same $g$ interval, showing that the avoided crossing redistributes two-qubit concurrence among the three nearby eigenbranches.
Figures~\ref{fig:triplet_groups}(c) and \ref{fig:triplet_groups}(d) show the corresponding finite-$\epsilon$ groups for $n=1$ and $n=2$.
Their concurrence panels, Figs.~\ref{fig:triplet_groups}(g) and \ref{fig:triplet_groups}(h), show the same redistribution pattern, with additional narrow changes caused by extra avoided crossings.

The exact Bell-basis network and its fixed-$n$ reduction in Eq.~\eqref{eq:Heff_n} share the three-component connectivity,
\begin{equation}
\ket{+}\leftrightarrow \ket{B_+}\leftrightarrow \ket{B_-}.
\end{equation}
In the displaced frame, the detuning supplies the $\ket{+}\leftrightarrow\ket{B_+}$ link, while the Franck-Condon-dressed qubit splitting supplies the $\ket{B_+}\leftrightarrow\ket{B_-}$ link.
The component $\ket{B_+}$ is therefore the intermediate state through which the $\ket{+}$ and $\ket{B_-}$ weights are exchanged.
Finite $\epsilon$ then allows the tracked eigenbranches to redistribute their Bell-basis weights, which changes their reduced two-qubit concurrence.
The higher-$n$ comparisons in Appendix~\ref{app:polaron} show that the same effective-chain picture captures the redistribution pattern in the exact $n=1$ and $n=2$ groups.

\subsection{Entanglement growth on the selected branch}

\begin{figure}[tbp]
\centering
\includegraphics[width=3.38in]{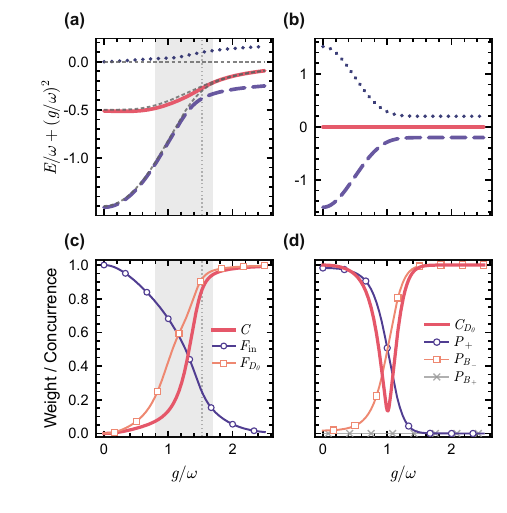}
\caption{
Fixed-$n=0$ reference and selected exact branch at $\Omega/\omega=0.75$ and $\epsilon/\omega=0.10$.
(a) Flattened exact spectrum, where the red solid curve denotes the selected branch, the gray dashed curves give the corresponding $\epsilon=0$ reference, and the shaded region marks the main avoided-crossing window.
(c) Concurrence $C$, weak-coupling overlap $F_{\rm in}$, and $D_0$-reference fidelity $F_{D_0}$ along the selected exact branch.
(b) Fixed-$n=0$ effective ladder.
(d) Basis weights $P_+$, $P_{B_-}$, $P_{B_+}$, and reduced-spin concurrence $C_{D_0}$ of the $D_0$ reference.
The colors and markers in the lower panels follow the in-panel legends, and the vertical dotted line marks the coupling at which the selected branch is closest in energy to any adjacent triplet branch.
The comparison shows that the exact branch leaves the weak-coupling state near the avoided crossing and approaches the full $D_0$ reference, while the effective model shows the transfer of weight from $\ket{+}$ to $\ket{B_-}$.
}
\label{fig:n0_mechanism}
\end{figure}

Figure~\ref{fig:n0_mechanism} uses the representative parameters $\Omega/\omega=0.75$ and $\epsilon/\omega=0.10$, for which the selected low-energy branch shows a clear growth of two-qubit entanglement.
The left column follows the exact branch labeled by its right endpoint $E^{\rm R}_{0,0}$.
In panel (a), the gray dashed $\epsilon=0$ reference identifies the crossing that is opened into the shaded avoided-crossing window at finite $\epsilon$.
At weak coupling, this branch approaches a separable $g=0$ eigenstate, and its concurrence is close to zero.
As $g$ is increased through the shaded avoided-crossing window, the branch moves away from the weak-coupling eigenstate and its concurrence increases.
On the larger-$g$ side, the branch acquires dominant $\ket{B_-}$ character and its concurrence approaches unity.

To follow how the exact branch changes as $g$ is increased, we monitor its overlap with two reference states.
The first is the overlap with the weak-coupling endpoint state,
\begin{equation}
F_{\rm in}(g)=
\left|\left\langle \psi_{\rm in}\middle|\psi(g)\right\rangle\right|^2,
\end{equation}
where $\ket{\psi_{\rm in}}$ is the $g=0$ state of the tracked exact branch.
The second is $F_{D_0}$, the fidelity with the $n=0$ reference $\ket{\mathcal D_0^{\rm full}(g)}$ defined in Eqs.~\eqref{eq:Dn_full} and \eqref{eq:Dn_fidelity_full}.
As shown in Fig.~\ref{fig:n0_mechanism}(c), $F_{\rm in}$ is close to unity at small $g$, while the concurrence is close to zero.
Near the shaded avoided-crossing window, $F_{\rm in}$ decreases and $F_{D_0}$ rises.
This exchange of overlaps shows that the exact branch moves away from the weak-coupling eigenstate and approaches the $D_0$ reference.

Panel (b) gives the corresponding fixed-$n=0$ effective ladder, and the right column then isolates the $D_0$ reference defined from Eq.~\eqref{eq:Dn}.
This effective reference has no $\ket{B_+}$ component and contains only $\ket{+}$ and $\ket{B_-}$.
With $G_0=e^{-2(g/\omega)^2}$ and $\Lambda_0=\sqrt{\epsilon^2+\Omega^2G_0^2}$, it can be written as
\begin{equation}
\ket{D_0(g)}
=
a(g)\ket{+}+b(g)\ket{B_-},
\end{equation}
where
\begin{equation}
a(g)=\frac{\Omega G_0}{\Lambda_0},
\qquad
b(g)=\frac{\epsilon}{\Lambda_0}.
\end{equation}
Its $\ket{B_-}$ weight is
\begin{equation}
\left|\left\langle B_-\middle|D_0(g)\right\rangle\right|^2
=
\frac{\epsilon^2}{\epsilon^2+\Omega^2e^{-4(g/\omega)^2}}.
\end{equation}
Since $G_0$ decreases as the two displaced oscillator packets separate, the $\ket{+}$ weight in Fig.~\ref{fig:n0_mechanism}(d) falls and the $\ket{B_-}$ weight rises, while the $\ket{B_+}$ weight remains zero.

When the oscillator part is included, the $\ket{+}$ and $\ket{B_-}$ components carry different displaced oscillator states.
The reduced-spin concurrence of this reference is
\begin{equation}
C_{D_0}(g)
=
\sqrt{(a^2-b^2)^2+4a^2b^2G_0^2}.
\end{equation}
This expression gives the red curve in Fig.~\ref{fig:n0_mechanism}(d).
The dip occurs when both Bell components are appreciable.
Their oscillator packets are then partly distinguishable, and tracing out the oscillator weakens the coherence between them.
At larger $g$, the $\ket{B_-}$ component dominates and the red curve returns to unity.

The concurrence growth in Fig.~\ref{fig:n0_mechanism} therefore occurs along an exact branch that starts from a separable weak-coupling eigenstate, passes through the avoided-crossing window, and develops a large overlap with the $D_0$ reference at larger $g$.
We next determine whether this $D_0$-reference branch has a high-concurrence or separable weak-coupling endpoint.

\subsection{Static entrance criterion}

For finite nonzero $\epsilon$, the relevant triplet crossings are lifted and the branch carrying $E^{\rm R}_{0,0}$ can be followed continuously between the two solvable limits.
The endpoint energies in Eqs.~\eqref{eq:g0_energies} and \eqref{eq:large_g_energies} reduce this branch connection to a level-ordering problem.
For $0<|\epsilon|<\omega/2$, $E^{\rm R}_{0,0}$ is the second triplet level in the large-coupling limit, and its weak-coupling endpoint is the lower of $E^{\rm L}_{0,0}$ and $E^{\rm L}_{1,-}$.
At $|\epsilon|=\omega/2$, $E^{\rm R}_{0,0}$ becomes degenerate with $E^{\rm R}_{1,-}$.
A connection to $E^{\rm L}_{0,0}$ gives a high-to-high concurrence branch, whereas a connection to $E^{\rm L}_{1,-}$ gives the low-to-high concurrence branch observed in Fig.~\ref{fig:n0_mechanism}.

The two weak-coupling energies are
\begin{equation}
E^{\rm L}_{0,0}=0
\label{eq:E00}
\end{equation}
and
\begin{equation}
E^{\rm L}_{1,-}=\omega-2\sqrt{\Omega^2+\epsilon^2}.
\label{eq:E1m}
\end{equation}
The energy $E^{\rm L}_{0,0}$ is degenerate in the full spin space at $g=0$, and the level itself does not define a unique concurrence.
Within the triplet block, the $D_0$ reference has the weak-coupling limit $\ket{D_0(0)}$, whose concurrence is unity.
By contrast, the weak-coupling eigenstate associated with $E^{\rm L}_{1,-}$, denoted by $\ket{\psi_{1,-}^{\rm L}}$, is a product-spin state at $g=0$ and has zero concurrence.

The low-to-high endpoint connection is obtained when the separable endpoint lies below the $D_0(0)$ endpoint,
\begin{equation}
E^{\rm L}_{1,-}<E^{\rm L}_{0,0}.
\label{eq:sorting_inversion}
\end{equation}
Substituting the weak-coupling energies gives the criterion
\begin{equation}
2\sqrt{\Omega^2+\epsilon^2}>\omega.
\label{eq:criterion}
\end{equation}
For $0<|\epsilon|<\omega/2$, Eq.~\eqref{eq:criterion} gives the endpoint ordering required for the minimal $n=0$ low-to-high connection.
Appendix~\ref{app:weak_ordering} shows where the $E^{\rm L}_{1,-}$ and $E^{\rm L}_{0,0}$ levels cross relative to the adjacent weak-coupling levels.

\begin{figure}[tbp]
\centering
\includegraphics[width=\linewidth]{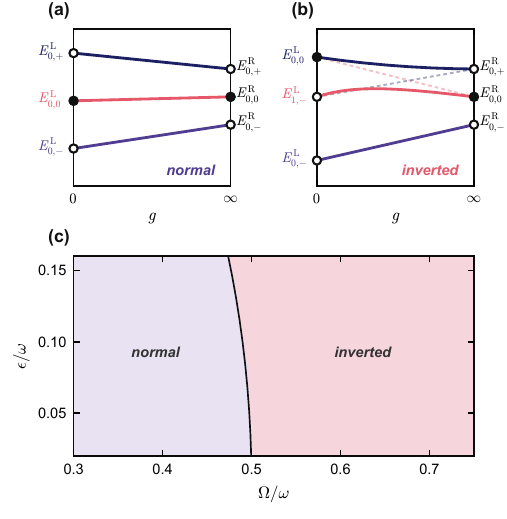}
\caption{
Static low-to-high concurrence route and its weak-coupling entrance criterion.
Panels (a) and (b) schematically compare the normal three-level weak-coupling ordering with the inverted ordering in which $E^{\rm L}_{1,-}$ drops below $E^{\rm L}_{0,0}$.
In both panels, the solid curves trace the branch connections between the weak- and strong-coupling endpoints.
In panel (b), the pale dashed lines show the uncoupled crossing underlying the avoided-crossing connection.
Open and filled endpoint markers denote reference endpoints with two-qubit concurrence $C=0$ and $C=1$, respectively.
Panel (c) tests the criterion by exact branch tracking in the $(\Omega/\omega,\epsilon/\omega)$ plane: the colors identify the weak-coupling endpoint reached by the branch carrying the right endpoint label $E^{\rm R}_{0,0}$.
The labels normal and inverted identify the static endpoint-ordering regimes; they do not indicate dynamical protocol success.
The black boundary is the analytic sorting criterion $2\sqrt{\Omega^2+\epsilon^2}=\omega$.
}
\label{fig:criterion_map}
\end{figure}

In the dimensionless $(\Omega/\omega,\epsilon/\omega)$ plane used in Fig.~\ref{fig:criterion_map}, Eq.~\eqref{eq:criterion} gives the boundary
\begin{equation}
\left(\frac{\Omega}{\omega}\right)^2+
\left(\frac{\epsilon}{\omega}\right)^2
=
\frac14.
\label{eq:arc}
\end{equation}
This boundary is specific to the minimal $n=0$ connection discussed above.
Other endpoint connections would involve different weak-coupling competitors and, for higher $n$ groups, may pass through several avoided-crossing windows.

Figure~\ref{fig:criterion_map} shows the static sorting criterion in schematic and exact-tracking forms.
Panels (a) and (b) compare the two possible weak-coupling orderings.
In the normal ordering, the branch carrying $E^{\rm R}_{0,0}$ connects back to the $D_0(0)$ endpoint at $g=0$, and both endpoints have high concurrence.
In the inverted ordering, $E^{\rm L}_{1,-}$ lies below $E^{\rm L}_{0,0}$, and the same branch connects instead to the separable endpoint.
Panel (c) tests this classification by exact triplet-spectrum tracking in the $(\Omega/\omega,\epsilon/\omega)$ plane.
The black curve is the analytic boundary in Eq.~\eqref{eq:criterion}.

For example, the point $(\Omega/\omega,\epsilon/\omega)=(0.40,0.03)$ lies on the criterion-fail side of the boundary,
\begin{equation*}
2\sqrt{\Omega^2+\epsilon^2}\approx 0.802\,\omega<\omega,
\end{equation*}
and the branch carrying $E^{\rm R}_{0,0}$ remains connected to $E^{\rm L}_{0,0}$ at weak coupling.
By contrast, the point $(\Omega/\omega,\epsilon/\omega)=(0.70,0.08)$ lies on the criterion-pass side,
\begin{equation*}
2\sqrt{\Omega^2+\epsilon^2}\approx 1.409\,\omega>\omega,
\end{equation*}
and the same branch connects instead to the separable endpoint $E^{\rm L}_{1,-}$.
Across the plotted parameter plane, Fig.~\ref{fig:criterion_map}(c) shows the same classification: the tracked branch connects to $E^{\rm L}_{0,0}$ on the criterion-fail side and to $E^{\rm L}_{1,-}$ on the criterion-pass side.
Thus, within the scanned low-energy window, the weak-coupling inversion gives the static sorting condition for the minimal $n=0$ connection.

\section{Bell-State Preparation}
\label{sec:bell}

\subsection{Preparation task and control schedules}

The ordering criterion identifies when the selected high-entanglement branch has a separable weak-coupling entrance.
We now determine the coupling and time needed to use this connection for state preparation.
For each parameter point with $0<\epsilon/\omega<1/2$ satisfying Eq.~\eqref{eq:criterion}, we initialize the system in $\ket{\psi_{1,-}^{\rm L}}$ at $g=0$.
We keep $\Omega$ and $\epsilon$ fixed and increase only $g$ to prepare a Bell-like state on the exact eigenbranch connecting this entrance to the large-coupling $D_0$ reference.
Its static coupling threshold, $g_f^{(0.95)}$, is the smallest coupling found at which the branch state's reduced-spin Bell fidelity reaches $0.95$.
This threshold depends on the eigenstate structure and is independent of the ramp shape.

To assess Bell-state preparation on the selected eigenbranch, we monitor both the reduced-spin Bell fidelity and the overlap with that branch.
The Bell fidelity measures how closely the two-qubit state approaches $\ket{B_-}$:
\begin{equation}
F_{B_-}(t)=\bra{B_-}\rho_{12}(t)\ket{B_-},
\end{equation}
where $\rho_{12}(t)$ is the reduced two-qubit density matrix.
To track occupation of the selected eigenbranch during the same evolution, we use
\begin{equation}
F_{\rm br}(t)=\left|\left\langle\psi_{\rm br}(g(t))\middle|\Psi(t)\right\rangle\right|^2,
\label{eq:branch_fidelity}
\end{equation}
where $\ket{\Psi(t)}$ is the evolved spin-oscillator state and $\ket{\psi_{\rm br}(g)}$ is the selected instantaneous eigenstate.
Preparation stops when $F_{B_-}$, $F_{\rm br}$, and the target eigenstate's own Bell fidelity first simultaneously reach $0.95$.
The three conditions require a Bell-like target branch and successful preparation near that branch at the stopping time.
We increase the coupling monotonically under the coupling and actual-time bounds,
\begin{equation}
0\leq g/\omega\leq2,\qquad \omega t\leq2000.
\label{eq:control_bounds}
\end{equation}

We compare a constant-speed linear ramp with a local adiabatic (LA) schedule constructed from the exact triplet spectrum.
The LA schedule allocates time using the coupling matrix elements divided by the squared energy gaps, with an adjustable lower bound on this weight.
Appendix~\ref{app:resource_convergence} gives the waveform construction, including how the ramp is smoothed at its endpoints.
We vary the overall duration of both ramps and, for LA, the time allocation along the ramp.
For each evolution, we locate the first time at which all three fidelity conditions are satisfied.
We denote the shortest actual stopping time found in this search by $T$.
The actual stopping coupling $g(T)$ can exceed $g_f^{(0.95)}$ without reaching the upper bound $2\omega$.

\subsection{Coupling thresholds and preparation times}

\begin{figure}[tbp]
\centering
\includegraphics[width=3.38in]{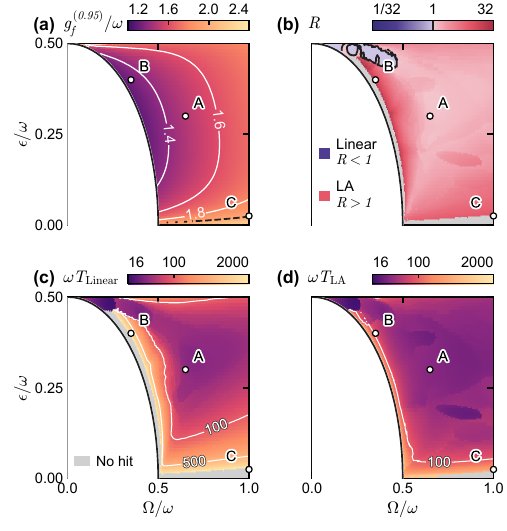}
\caption{Static coupling threshold and preparation time.
(a) Minimum coupling $g_f^{(0.95)}$ at which the target eigenstate reaches a Bell fidelity of $0.95$; the dashed contour marks $g_f^{(0.95)}/\omega=2$.
(b) Time ratio $R=T_{\rm Linear}/T_{\rm LA}$ where both protocols succeed: $R<1$ favors Linear and $R>1$ favors LA; the thick black contour marks $R=1$.
(c),(d) Linear and LA preparation times, respectively.
Here $T$ is the shortest actual stopping time found for which the target eigenstate's Bell fidelity, $F_{B_-}$, and $F_{\rm br}$ all reach $0.95$.
The black curve $2\sqrt{\Omega^2+\epsilon^2}=\omega$ bounds the white region where the selected entrance route is absent.
Light gray (``No hit'') in (c),(d) denotes $g_f^{(0.95)}>2\omega$ or no joint success found by $\omega t=2000$ for the respective protocol; in (b), it denotes an undefined ratio because either protocol lacks a successful result.
A, B, and C mark $(\Omega/\omega,\epsilon/\omega)=(0.65,0.30),(0.35,0.40),(1,0.025)$.}
\label{fig:adiabatic_generation}
\end{figure}

Figure~\ref{fig:adiabatic_generation} shows the static coupling threshold and the Linear and LA preparation times across the $(\Omega/\omega,\epsilon/\omega)$ plane.
The black curve in each panel marks the boundary of the region satisfying the analytic entrance criterion.
Within this region, panel (a) gives $g_f^{(0.95)}$.
Values above $2\omega$ identify parameters for which the target eigenstate cannot reach the required Bell fidelity within this bound.
Panels (c) and (d) show the Linear and LA preparation times, respectively, on the same color scale.
Panel (b) shows the ratio $R=T_{\rm Linear}/T_{\rm LA}$ where both methods succeed.
Values $R>1$ indicate faster LA preparation, whereas $R<1$ indicates faster Linear preparation.
Numerical checks are reported in Appendix~\ref{app:resource_convergence}.

Local adiabatic scheduling reduces the preparation time over a broad part of the accessible region and extends the region of success within the common budget.
The additional successful points arise from the changed dynamics: the target eigenstates and their static coupling thresholds are identical for the two schedules.
The gain varies substantially across the map.
At A, LA gives a modest reduction, from $\omega T=34.28$ to $28.92$.
At B, LA reduces the time from $1101.45$ to $59.93$, a factor of $18.38$.
At C, LA succeeds at $\omega T=319.45$, whereas the Linear search finds no successful preparation by $\omega t=2000$.
Among the 4778 sampled points where both methods succeed, LA is faster at 4553 and Linear at 225; 661 further points show LA-only success.
These targeted sample counts are not area fractions in the parameter plane.

\subsection{Preparation dynamics and accuracy}

\begin{figure}[tbp]
\centering
\includegraphics[width=3.38in]{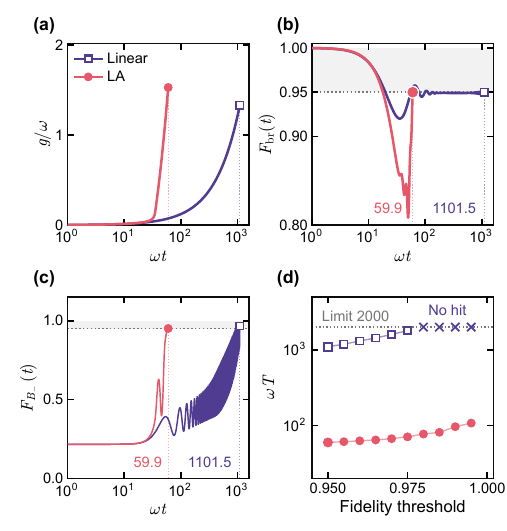}
\caption{Control dynamics and preparation precision at point B, $(\Omega/\omega,\epsilon/\omega)=(0.35,0.40)$.
(a) Coupling, (b) target-branch fidelity, and (c) Bell fidelity.
Endpoint markers and vertical guides mark the first stops satisfying all three $0.95$ fidelity conditions.
The light-gray bands in (b,c) show where the displayed fidelity is at least $0.95$, with dotted lines at the threshold.
(d) Preparation time as all three thresholds are raised together; faint connecting lines guide the eye.
Purple crosses on the ``Limit 2000'' line mark thresholds for which the Linear search found no joint success by $\omega t=2000$.}
\label{fig:trajectory_diagnostics}
\end{figure}

At point B, LA spends most of its preparation time near the weak-coupling entrance and accelerates at larger coupling, whereas Linear maintains the same slow rate throughout the ramp [Fig.~\ref{fig:trajectory_diagnostics}(a)].
The faster sweep reaches a larger stopping coupling in much less time.
During this shorter evolution, the branch fidelity falls to about $0.81$ before recovering to $0.95$ [panel (b)].
For Linear, the Bell fidelity continues to oscillate after the branch fidelity has nearly stabilized [panel (c)], reflecting coherent admixtures of other eigenstates.
For LA, the Bell fidelity reaches $0.95$ before the branch fidelity recovers; the return to the target branch therefore determines the joint stopping time.

The $0.95$ threshold provides a common reference for the maps; higher thresholds test the attainable preparation accuracy.
At point B, Fig.~\ref{fig:trajectory_diagnostics}(d) shows LA preparation times $\omega T=70.99$, $96.66$, and $107.86$ for joint thresholds $0.975$, $0.99$, and $0.995$, respectively.
Linear reaches $0.975$ at $\omega T=1797.79$; its searches at $0.980$, $0.985$, $0.990$, and $0.995$ find no joint hit within $\omega t\leq2000$.
LA therefore retains its time advantage at B as preparation accuracy increases.
At point C, the target eigenstate's Bell fidelity is at most about $0.9569$ within $g/\omega\leq2$, ruling out higher joint thresholds for both schedules.
Schedule design controls how quickly the target can be prepared, while the target eigenstate sets an upper bound on the joint threshold within the available coupling range.

A quantitative comparison with other Bell-state preparation protocols is given in Appendix~\ref{app:protocol_comparison}.

\section{Discussion}
\label{sec:discussion}

\subsection{Experimental relevance}

The model studied in this work is relevant to platforms in which two qubits couple to a shared motional mode, with trapped Rydberg ions providing a natural example~\cite{Hamlyn2024}.
In such a realization, the collective motional mode plays the role of the oscillator $a$, and laser-controlled spin-motion processes engineer the two-spin oscillator coupling $g(a+a^\dagger)\sigma_1^x\sigma_2^x$.
The effective parameters $\Omega$ and $\epsilon$ are set by the corresponding driving and detuning controls.
Related trapped-ion work has demonstrated tunable QRM simulation and QRM critical behavior, while trapped-Rydberg-ion experiments and reviews establish coherent Rydberg excitation, confinement, and control in ion traps~\cite{Pedernales2015TrappedIonQRM,Lv2018TrappedIonQRM,Cai2021QRMQPT,Higgins2017RydbergIonPRX,Higgins2017RydbergIonPRL,Mokhberi2020RydbergIons}.

For a concrete parameter scale, we consider the microwave-dressed $n=80$ Rydberg states of two $^{88}\mathrm{Sr}^{+}$ ions in the proposal of Hamlyn et al.~\cite{Hamlyn2024}.
The axial trap frequency $\nu/(2\pi)=2.02\,\mathrm{MHz}$ gives a breathing-mode frequency $\omega/(2\pi)=\sqrt{3}\nu/(2\pi)\simeq3.50\,\mathrm{MHz}$.
Retaining $\Omega/(2\pi)=25\,\mathrm{kHz}$ from their estimate, the reported ratio $g/g_c\simeq1.04$, with $g_c=\sqrt{\omega\Omega}$, corresponds to $g/(2\pi)\simeq0.31\,\mathrm{MHz}$.

At this oscillator frequency, the representative LA protocol B requires $\Omega/(2\pi)\simeq1.22\,\mathrm{MHz}$ and $\epsilon/(2\pi)\simeq1.40\,\mathrm{MHz}$.
For a joint fidelity threshold of $0.95$, its stop at $g(T)/\omega=1.530$ and duration $\omega T=59.93$ correspond to $g(T)/(2\pi)\simeq5.36\,\mathrm{MHz}$ and $T\simeq2.73\,\mu\mathrm{s}$.
The required stopping coupling is about seventeen times that of the reference example.
Implementing this waveform therefore requires the corresponding increase in the collective coupling and coherent spin-motion evolution over several microseconds.

Appendix~\ref{app:noise} examines effective-bias dephasing, motional heating, and Rydberg loss out of the qubit encoding for the LA waveform at B that reaches the coherent joint threshold $0.99$ in Fig.~\ref{fig:trajectory_diagnostics}(d).

\subsection{Scope and possible generalizations}

The quantitative Bell-state preparation analysis in this work is restricted to the minimal $n=0$ branch connection.
For this route, the weak-coupling inversion $E^{\rm L}_{1,-}<E^{\rm L}_{0,0}$ identifies the exact branch that starts from the separable state $\ket{\psi_{1,-}^{\rm L}}$.
As $g$ is increased, this branch passes through the finite-$\epsilon$ avoided-crossing window and approaches the lifted $D_0$ reference at large coupling.
The inequality $2\sqrt{\Omega^2+\epsilon^2}>\omega$ is the corresponding static endpoint-sorting condition.
The static eigenstate sets the threshold coupling, while the ramp shape and its speed determine the finite-time preparation requirement.

The same endpoint comparison can be made for weak-coupling states with $\ell$ oscillator quanta on the lower dressed-spin branch.
For the weak-coupling level $E^{\rm L}_{\ell,-}$ with $\ell=1,2,\ldots$, the energy is
\begin{equation}
E^{\rm L}_{\ell,-}=\ell\omega-2\sqrt{\Omega^2+\epsilon^2}.
\label{eq:Em}
\end{equation}
This level crosses $E^{\rm L}_{0,0}$ at
\begin{equation}
2\sqrt{\Omega^2+\epsilon^2}=\ell\omega.
\label{eq:criterion_m}
\end{equation}
It lies below $E^{\rm L}_{0,0}$ when $2\sqrt{\Omega^2+\epsilon^2}>\ell\omega$.
For $\ell=1$, this gives the minimal route analyzed in the main text.
For $\ell>1$, the same comparison marks when a separable weak-coupling endpoint with $\ell$ additional oscillator quanta is reordered below $E^{\rm L}_{0,0}$.
For a branch connected at large coupling to a higher $D_n$ reference, the analogous weak-coupling comparison is between $E^{\rm L}_{n+\ell,-}$ and $E^{\rm L}_{n,0}$, giving the same threshold $2\sqrt{\Omega^2+\epsilon^2}=\ell\omega$.
Such higher-order paths may pass through several finite-$\epsilon$ avoided-crossing windows.
These pairwise comparisons identify candidate entrance levels.
The subsequent avoided crossings determine how each candidate level is carried through the spectrum toward the large-$g$ $D_n$ reference.
A complete preparation analysis must then specify the exact branch connection, the relevant gaps, and the finite-time ramp response.

\section{Conclusion}
\label{sec:conclusion}

In this work, we have studied two-qubit entanglement in the tripartite quantum Rabi model (TQRM), focusing on how the collective spin-oscillator coupling reorganizes the spectrum and supports Bell-state preparation.
At $\epsilon=0$, the triplet spectrum separates into a zero-concurrence QRM sector and a displaced-oscillator Bell sector.
Finite detuning couples these two triplet structures and turns their crossings into avoided crossings, where the eigenstate concurrence is redistributed.
In the low-energy spectrum, this mixing produces an exact branch that starts from the separable weak-coupling state $\ket{\psi_{1,-}^{\rm L}}$, passes through the avoided-crossing window, and approaches the lifted $D_0$ reference at large coupling.
Along this branch, the reduced spin state approaches a $\ket{B_-}$-like endpoint.
The three-state effective model explains how the $\ket{B_-}$ component becomes dominant as the coupling increases.
For the minimal route analyzed here, the ordering of $E^{\rm L}_{1,-}$ and $E^{\rm L}_{0,0}$ at weak coupling gives the static criterion for whether the selected exact branch has a separable endpoint.

After identifying this static branch connection, we tested it with finite-time linear and local adiabatic ramps under common coupling and time bounds.
The protocol-resource maps identify the static threshold coupling along this branch and the earliest preparation times found under the joint fidelity requirement.
Local adiabatic scheduling reduces the preparation time over a broad part of the accessible parameter region and extends the region of successful preparation within the common resource bounds.
Together, the static spectrum and the ramp benchmarks quantify the coupling and time needed to reach high Bell-state fidelity along the selected branch.
Future work can extend this branch-based framework to higher-order connections, broader control families~\cite{Chen2021STAQRM,Shu2016FrequencyControl}, and device-specific noise models.

\begin{acknowledgments}
This work was supported by the National Natural Science Foundation of China (Grant No. 12205383, 12274470) and the Hunan Provincial Natural Science Foundation (Grant No. 2024JJ6483).
\end{acknowledgments}

\appendix
\renewcommand{\topfraction}{0.95}
\renewcommand{\bottomfraction}{0.85}
\renewcommand{\textfraction}{0.05}
\renewcommand{\floatpagefraction}{0.80}

\section{Bell-basis representation and singlet decoupling}
\label{app:bell_basis}

In this appendix we collect the spin algebra underlying the Bell-basis representation used in Sec.~\ref{sec:model}.
Starting from the two-qubit $z$ basis in the display order used in Fig.~\ref{fig:model_overview}(a),
\begin{equation}
\ket{\uparrow\uparrow},\qquad
\ket{\downarrow\downarrow},\qquad
\ket{\uparrow\downarrow},\qquad
\ket{\downarrow\uparrow},
\end{equation}
we define the Bell-adapted states
\begin{equation}
\ket{B_+}=\frac{\ket{\uparrow\uparrow}+\ket{\downarrow\downarrow}}{\sqrt2},
\qquad
\ket{B_-}=\frac{\ket{\uparrow\uparrow}-\ket{\downarrow\downarrow}}{\sqrt2},
\end{equation}
and
\begin{equation}
\ket{+}=\frac{\ket{\uparrow\downarrow}+\ket{\downarrow\uparrow}}{\sqrt2},
\qquad
\ket{-}=\frac{\ket{\uparrow\downarrow}-\ket{\downarrow\uparrow}}{\sqrt2}.
\end{equation}
Using the ordered basis
\begin{equation}
\mathcal B=\{\ket{B_+},\ \ket{+},\ \ket{B_-},\ \ket{-}\},
\end{equation}
the three spin operators relevant to the TQRM take the forms
\begin{equation}
\sigma_1^x\sigma_2^x=
\begin{pmatrix}
1&0&0&0\\
0&1&0&0\\
0&0&-1&0\\
0&0&0&-1
\end{pmatrix},
\end{equation}
\begin{equation}
\sigma_1^x+\sigma_2^x=
\begin{pmatrix}
0&2&0&0\\
2&0&0&0\\
0&0&0&0\\
0&0&0&0
\end{pmatrix},
\end{equation}
and
\begin{equation}
\sigma_1^z+\sigma_2^z=
\begin{pmatrix}
0&0&2&0\\
0&0&0&0\\
2&0&0&0\\
0&0&0&0
\end{pmatrix}.
\end{equation}
These matrices show that $\sigma_1^x\sigma_2^x$ separates the Bell basis into two displacement classes with eigenvalues $\eta=\pm1$.
The states $\ket{B_+}$ and $\ket{+}$ have $\eta=+1$,
\begin{equation}
\sigma_1^x\sigma_2^x\ket{B_+}=\ket{B_+},\qquad
\sigma_1^x\sigma_2^x\ket{+}=\ket{+},
\end{equation}
whereas $\ket{B_-}$ and $\ket{-}$ have $\eta=-1$,
\begin{equation}
\sigma_1^x\sigma_2^x\ket{B_-}=-\ket{B_-},\qquad
\sigma_1^x\sigma_2^x\ket{-}=-\ket{-}.
\end{equation}
At the same time, the detuning-derived single-spin term mixes only $\ket{B_+}$ and $\ket{+}$, while the qubit-splitting term mixes only $\ket{B_+}$ and $\ket{B_-}$.

In the ordered spin basis $\mathcal B$, the Hamiltonian takes the block form
\begin{equation}
\resizebox{0.96\columnwidth}{!}{$
\displaystyle
H=
\omega a^\dagger a\,\mathbb I_4
+
\begin{pmatrix}
g(a+a^\dagger) & 2\epsilon & -2\Omega & 0\\
2\epsilon & g(a+a^\dagger) & 0 & 0\\
-2\Omega & 0 & -g(a+a^\dagger) & 0\\
0 & 0 & 0 & -g(a+a^\dagger)
\end{pmatrix}.
$}
\label{eq:H_bell_appendix}
\end{equation}

\begin{figure}[!tbp]
\centering
\includegraphics[width=\linewidth]{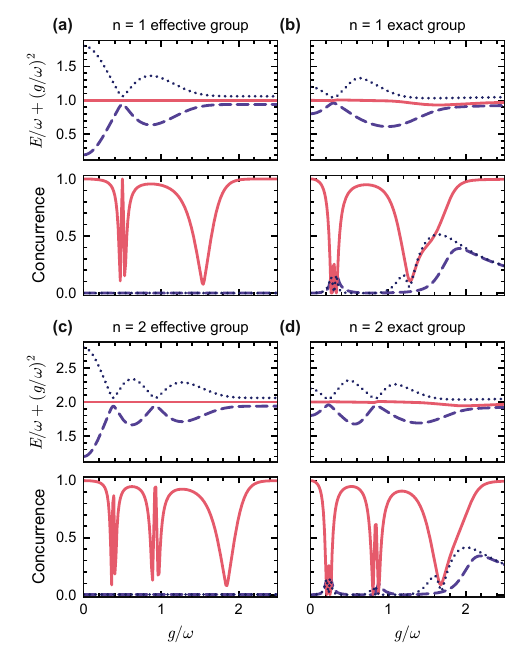}
\caption{
Effective groups and corresponding exact tracked groups for the higher triplet branches at the finite-$\epsilon$ parameters of Fig.~\ref{fig:triplet_groups}, $\Omega/\omega=0.40$ and $\epsilon/\omega=0.03$.
Each labeled pair contains the flattened energies above and the concurrences of the same three branches below, with the same colors and line styles used in both rows.
Purple dashed, red solid, and dark-purple dotted curves denote the $-$, $0$, and $+$ branches, respectively.
Panels (a,b) compare the $n=1$ effective group with the corresponding exact group.
Panels (c,d) give the analogous comparison for the $n=2$ groups.
Within each pair, the effective and exact energy panels share the same vertical scale.
As in Fig.~\ref{fig:triplet_groups}, the exact groups are labeled by the large-$g$ displaced-oscillator limits of the tracked branches.
}
\label{fig:higher_n_effective_exact}
\end{figure}

Equation~\eqref{eq:H_bell_appendix} is exactly block diagonal: the triplet block is spanned by the first three basis vectors, while the singlet state $\ket{-}$ forms an invariant one-dimensional block governed by
\begin{equation}
H_-=\omega a^\dagger a-g(a+a^\dagger),
\end{equation}
Its eigenstates are
\begin{equation}
\ket{-}\otimes D(\alpha)\ket{n},
\qquad
\alpha=\frac{g}{\omega},
\end{equation}
with displaced-oscillator energies
\begin{equation}
E_{-,n}=n\omega-\frac{g^2}{\omega}.
\end{equation}
Here $D(\alpha)=\exp[\alpha(a^\dagger-a)]$, and $\ket{n}$ is the oscillator Fock state.
This exact singlet decoupling is the algebraic origin of the triplet-only reorganization analyzed in the main text.

\section{Polaron transformation and fixed-\texorpdfstring{$n$}{n} effective triplet model}
\label{app:polaron}

This appendix derives the fixed-$n$ triplet model used in Secs.~\ref{sec:model} and \ref{sec:static}.
The starting point is the exact triplet block.
A spin-dependent polaron transformation absorbs the tripartite displacement term into a common oscillator energy, leaving the qubit-splitting term to carry Franck-Condon overlaps between displaced oscillator states.
Keeping the diagonal overlap $G_n(\alpha)$ gives the fixed-$n$ effective Hamiltonian, while the off-diagonal overlaps $G_{mn}(\alpha)$ with $m\ne n$ remain in the exact transformed-frame formulation.

In the triplet basis
\begin{equation}
\{\ket{B_+},\ \ket{+},\ \ket{B_-}\},
\end{equation}
the exact triplet block reads
\begin{equation}
H_{\rm triplet}
=
\omega a^\dagger a\,\mathbb I_3
+
\begin{pmatrix}
g(a+a^\dagger) & 2\epsilon & -2\Omega\\
2\epsilon & g(a+a^\dagger) & 0\\
-2\Omega & 0 & -g(a+a^\dagger)
\end{pmatrix}.
\label{eq:H_triplet_appendix}
\end{equation}
Introducing
\begin{equation}
X=\ket{B_+}\bra{B_+}+\ket{+}\bra{+}-\ket{B_-}\bra{B_-},
\qquad
X^2=\mathbb I_3,
\end{equation}
and the unitary transformation
\begin{equation}
U=e^{-\alpha(a^\dagger-a)X},
\qquad
\alpha=\frac{g}{\omega},
\end{equation}
one obtains
\begin{equation}
U^\dagger a\,U=a-\alpha X,
\qquad
U^\dagger a^\dagger U=a^\dagger-\alpha X.
\end{equation}

As a result,
\begin{equation}
U^\dagger\!\left[\omega a^\dagger a+g(a+a^\dagger)X\right]\!U
=
\omega a^\dagger a-\frac{g^2}{\omega}.
\end{equation}
Thus the tripartite coupling is absorbed into a constant energy shift in the transformed frame.

The remaining two terms behave differently.
Because $\ket{B_+}$ and $\ket{+}$ have the same $X=+1$ eigenvalue, the detuning-derived single-spin term is unchanged,
\begin{equation}
H_\epsilon=2\epsilon\bigl(\ket{B_+}\bra{+}+\ket{+}\bra{B_+}\bigr),
\qquad
U^\dagger H_\epsilon U=H_\epsilon.
\end{equation}
By contrast, $\ket{B_+}$ and $\ket{B_-}$ have opposite $X$ eigenvalues, and the qubit-splitting term therefore acquires displacement operators,
\begin{equation}
\begin{aligned}
&U^\dagger[-2\Omega(\ket{B_+}\bra{B_-}+\ket{B_-}\bra{B_+})]U\\
&\quad =
-2\Omega\,\ket{B_+}\bra{B_-}D(2\alpha)
-2\Omega\,\ket{B_-}\bra{B_+}D(-2\alpha).
\end{aligned}
\end{equation}
The exact transformed triplet Hamiltonian is therefore
\begin{equation}
\begin{aligned}
\widetilde H_{\rm triplet}
&=
\left(\omega a^\dagger a-\frac{g^2}{\omega}\right)\mathbb I_3\\
&\quad
+2\epsilon\bigl(\ket{B_+}\bra{+}+\ket{+}\bra{B_+}\bigr)\\
&\quad
-2\Omega\,\ket{B_+}\bra{B_-}D(2\alpha)\\
&\quad
-2\Omega\,\ket{B_-}\bra{B_+}D(-2\alpha).
\end{aligned}
\label{eq:H_triplet_polaron}
\end{equation}

In the oscillator Fock basis of the transformed frame, the coupling between opposite $X$ sectors is encoded in the Franck-Condon matrix elements
\begin{equation}
G_{mn}(\alpha)=\bra{m}D(2\alpha)\ket{n}.
\end{equation}

The fixed-$n$ projection retains the diagonal matrix element
\begin{equation}
G_n(\alpha)\equiv G_{nn}(\alpha)
=
\bra{n}D(2\alpha)\ket{n}
=
e^{-2\alpha^2}L_n(4\alpha^2),
\end{equation}
and, for real $\alpha$, $\bra{n}D(-2\alpha)\ket{n}=G_n(\alpha)$ as well.

Projecting Eq.~\eqref{eq:H_triplet_polaron} onto the three spin components with the same transformed-frame Fock index $n$ therefore gives the fixed-$n$ effective Hamiltonian
\begin{equation}
H_n^{\rm eff}
=
\left(n\omega-\frac{g^2}{\omega}\right)\mathbb I
+
2
\begin{pmatrix}
0&\epsilon&-\Omega G_n(\alpha)\\
\epsilon&0&0\\
-\Omega G_n(\alpha)&0&0
\end{pmatrix},
\end{equation}

which is the effective model used in the main text.
Its three eigenvalues are
\begin{equation}
E_{n,0}^{\rm eff}=n\omega-\frac{g^2}{\omega},
\qquad
E_{n,\pm}^{\rm eff}=n\omega-\frac{g^2}{\omega}\pm 2\Lambda_n(g),
\end{equation}
with
\begin{equation}
\Lambda_n(g)=\sqrt{\epsilon^2+\Omega^2G_n(\alpha)^2}.
\end{equation}

The corresponding fixed-$n$ reference state is
\begin{equation}
\ket{D_n(g)}=
\frac{\Omega G_n(\alpha)\,\ket{+}+\epsilon\,\ket{B_-}}
{\sqrt{\epsilon^2+\Omega^2G_n(\alpha)^2}},
\end{equation}
which is the effective $D_n$ reference used in the main text.
Within the fixed-$n$ effective Hamiltonian, this spin state belongs to the $E_{n,0}^{\rm eff}$ branch because its net coupling to $\ket{B_+}$ cancels.
The two contributions to the $\ket{B_+}$ matrix element are $\epsilon\,\Omega G_n(\alpha)$ and $-\Omega G_n(\alpha)\epsilon$, whose sum is zero, leaving only the common displaced-oscillator energy.

The remaining two normalized eigenstates of the fixed-$n$ block are
\begin{equation}
\ket{\phi_{n,\pm}^{\rm eff}}
=\frac{1}{\sqrt{2}}\left[
\ket{B_+}\pm
\frac{\epsilon\ket{+}-\Omega G_n(\alpha)\ket{B_-}}{\Lambda_n(g)}
\right],
\label{eq:effective_vectors}
\end{equation}
with energies $E_{n,\pm}^{\rm eff}$.
Together with $\ket{D_n}$, they form an orthonormal eigenbasis whenever $\Lambda_n>0$.
At $\Lambda_n=0$, the block is proportional to the identity and any orthonormal triplet basis may be chosen.
To compare any of these states with a full eigenstate, its $\ket{B_+}$ and $\ket{+}$ components are multiplied by $D(-\alpha)\ket{n}$ and its $\ket{B_-}$ component by $D(+\alpha)\ket{n}$, as in Eq.~\eqref{eq:Dn_full}.

\begin{figure}[t]
\centering
\includegraphics[width=\linewidth]{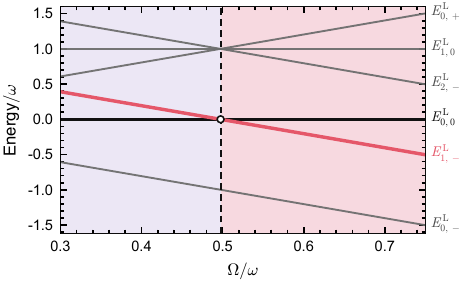}%
\vphantom{\includegraphics[width=\linewidth]{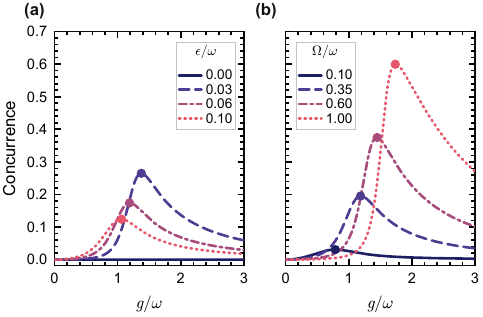}}
\caption{
Analytic weak-coupling endpoint ordering at $\epsilon/\omega=0.05$.
The black, red, and gray lines show $E^{\rm L}_{0,0}=0$, $E^{\rm L}_{1,-}=\omega-2\sqrt{\Omega^2+\epsilon^2}$, and the neighboring levels $E^{\rm L}_{0,\pm}$, $E^{\rm L}_{1,0}$, and $E^{\rm L}_{2,-}$, respectively.
The purple and rose backgrounds follow the normal and inverted regions in Fig.~\ref{fig:criterion_map}.
The black vertical dashed line and open circle mark $2\sqrt{\Omega^2+\epsilon^2}=\omega$, where $E^{\rm L}_{1,-}=E^{\rm L}_{0,0}$.
}
\label{fig:weak_ordering_appendix}
\end{figure}

Figure~\ref{fig:higher_n_effective_exact} compares the fixed-$n$ effective groups with the corresponding exact tracked groups for $n=1$ and $n=2$ at the parameters of Fig.~\ref{fig:triplet_groups}.
Panels (a) and (c) show the smoother fixed-$n$ reference trends.
The concurrence dips of the red solid $0$ branch reflect changes in the relative weights and oscillator overlap of its two oppositely displaced Bell components.
At a zero of $G_n(\alpha)$, the lifted reference state reduces to $\ket{B_-}\otimes D(+\alpha)\ket{n}$ and its concurrence returns to unity.
Panels (b) and (d) show the corresponding exact tracked branches.
The exact spectra retain the same large-scale concurrence pattern, with additional avoided crossings producing narrow dips in the red solid branch and visible concurrence transfer to the purple dashed and dark-purple dotted branches.
At larger $g/\omega$, the red solid concurrence recovers in both descriptions, showing that the exact branch remains organized by the fixed-$n$ reference trend.

The approximation omits matrix elements proportional to $2\Omega G_{mn}$ with $m\ne n$.
Its accuracy depends on these couplings relative to the energy separations of the groups they connect.
Near intergroup resonances, their mixing must be retained even when a diagonal overlap $G_n$ is small.
We therefore test the state and spectral approximations separately against the full triplet Hamiltonian.

We benchmark the $n=0$ reference on 466 parameter pairs selected from $0\leq\Omega/\omega\leq1$ and $0.025\leq\epsilon/\omega\leq0.475$, both in steps of $0.025$, by the condition $2\sqrt{\Omega^2+\epsilon^2}>\omega$.
For each pair we compare eight couplings $g/\omega=1.2,1.5,1.8,2,2.2,2.5,3,3.5$.
The selected branch is the second-lowest exact triplet eigenstate.
Its minimum $F_{D_0}$ over these pairs is $0.0989$, $0.9517$, and $0.9913$ at $g/\omega=1.2,2,2.5$, respectively.
At $g/\omega=2$ and $2.5$, the largest absolute error in its reduced Bell fidelity relative to the lifted $D_0$ state is $0.0430$ and $0.0075$, respectively.
Thus the target-state approximation becomes quantitatively useful on the sampled large-coupling domain.

For the complete three-state spectrum, we match the lifted effective eigenstates one-to-one to full triplet eigenstates by maximizing their total squared overlap.
At $g/\omega=2.5$, the median of the smallest of these three overlaps is $0.9933$; the smallest overlap across all parameter pairs is $0.5986$.
The largest matched energy error is $0.1674\omega$.
At point B and $g/\omega=2$, for example, $F_{D_0}\approx0.9975$, whereas the nearest exact triplet gap is approximately $0.215\omega$, compared with $2\Lambda_0\approx0.80\omega$ within the fixed-$n$ block: an eigenstate from another reference group lies between its matched levels.
The full gaps and matrix elements, rather than the projected gap, are consequently used to construct the LA schedule.
These tests, together with 497 comparisons along the representative and Fig.~\ref{fig:triplet_groups} parameter paths, were repeated with 48 and 64 oscillator states; the largest change in the compared overlaps, Bell fidelities, or energies (in units of $\omega$) is below $4.4\times10^{-9}$.
Together with Fig.~\ref{fig:higher_n_effective_exact}, these results show that the fixed-$n$ model captures the overall concurrence trends and provides an accurate reference for the target state at large coupling.

\section{Weak-coupling endpoint ordering}
\label{app:weak_ordering}

For the minimal $n=0$ connection, the main-text sorting condition involves the weak-coupling endpoints
\begin{equation}
E^{\rm L}_{0,0}=0,
\qquad
E^{\rm L}_{1,-}=\omega-2\sqrt{\Omega^2+\epsilon^2}.
\end{equation}
They exchange order at
\begin{equation}
2\sqrt{\Omega^2+\epsilon^2}=\omega.
\end{equation}
At fixed $\epsilon$, increasing $\Omega$ crosses this point and places $E^{\rm L}_{1,-}$ below $E^{\rm L}_{0,0}$.

In Fig.~\ref{fig:weak_ordering_appendix}, the black $E^{\rm L}_{0,0}$ line and the red $E^{\rm L}_{1,-}$ line show this exchange, while the gray curves give the neighboring weak-coupling levels around the selected crossing.

\vspace{0.5\baselineskip}
\section{Parameter dependence of static reorganization}
\label{app:param_response}

\begin{figure}[t]
\centering
\includegraphics[width=\linewidth]{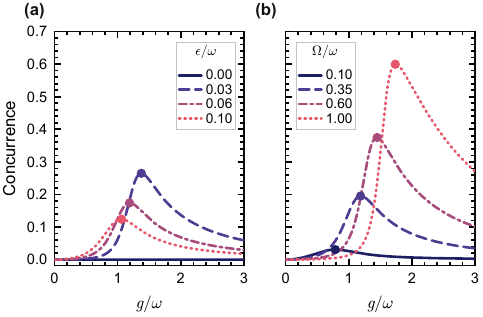}
\caption{
Parameter dependence of the static concurrence reorganization in the triplet-subspace ground state.
(a) Scan of $\epsilon/\omega$ at fixed $\Omega/\omega=0.25$.
(b) Scan of $\Omega/\omega$ at fixed $\epsilon/\omega=0.10$.
Filled circles indicate the nonzero concurrence peaks.
}
\label{fig:parameter_response}
\end{figure}

The avoided-crossing windows discussed in Sec.~\ref{sec:static} move and change shape when the control parameters are varied.
Figure~\ref{fig:parameter_response} illustrates this dependence through the triplet-subspace ground-state concurrence.
In panel (a), finite $\epsilon$ turns the nearly flat $\epsilon=0$ curve into a concurrence peak, which grows and then weakens as the detuning is increased.
In panel (b), increasing $\Omega$ shifts the peak toward larger $g/\omega$ and raises its maximum over the range shown.

\section{Control construction and numerical checks}
\label{app:resource_convergence}

\subsection{Local adiabatic schedules and stopping-time search}

Using exact instantaneous triplet eigenstates and energies, we define the local adiabatic indicator for the target branch of Sec.~\ref{sec:bell} as
\begin{equation}
\mathcal A(g)=\max_{m\ne{\rm br}}
\frac{|\bra{\psi_m(g)}\partial_g H\ket{\psi_{\rm br}(g)}|}
{|E_m(g)-E_{\rm br}(g)|^2}.
\label{eq:la_indicator}
\end{equation}
where $m$ runs over the other retained triplet eigenstates.
We interpolate $\mathcal A(g)$ along the target branch and let $\tau$ denote the nominal duration of a complete ramp to $g=2\omega$.
The linear family is $g(t)=2\omega t/\tau$.
For LA, we use $g(\theta)=\omega(1-\cos\theta)$, $0\leq\theta\leq\pi$, with
\begin{equation}
\frac{dt}{d\theta}
=\lambda\max[\mathcal A(g(\theta)),\mathcal A_0]\omega\sin\theta
+\frac{0.04\tau}{\pi}.
\label{eq:la_schedule}
\end{equation}
The nonnegative floor $\mathcal A_0$ changes the time allocation: $\mathcal A_0=0$ weights the ramp by the local adiabatic indicator, while increasing the floor assigns more time to the regions with smaller $\mathcal A$.
We vary $\mathcal A_0$ and $\tau$, retaining the fastest validated waveform as the search is refined.
The positive constant $\lambda$ makes the integral equal to $\tau$; the last term regularizes the endpoints of the complete waveform and gives zero endpoint slope.
A cubic Hermite interpolation yields the waveform, whose coupling range and monotonicity are checked directly.
Preparation uses only the prefix ending at the first joint threshold crossing.
Thus $\tau$ is a speed parameter and may exceed the time budget while the actual stop satisfies $\omega T\leq2000$.
After stopping, a fixed Hamiltonian conserves the target-eigenbranch population, while coherent admixtures can still produce Bell-fidelity oscillations.

For the new searches at a joint threshold $\eta$, the short-duration scan starts at $\omega\tau=0.99[\sqrt{\eta}-\sqrt{F_{B_-}(0)}]/(2\sqrt{3})$, just below a conservative lower bound on the preparation time.
The upper bound is determined separately for each point and waveform: let $s_-$ be the normalized time $t/\tau$ at the coupling two spectral nodes before the first sampled target-state Bell fidelity reaching $\eta$.
With no successful candidate, the nominal upper bound is $\omega\tau=2000/s_-$; after finding a stop at $T_{\rm best}$, it becomes $\omega\tau=\min(2000,\omega T_{\rm best}+0.02)/s_-$, retaining a margin for event localization.
Previously computed same-point long-duration Linear searches are reused where available, and the added short-duration searches cover $\omega\tau\leq40$.
At spatial points without such history, and for the higher thresholds at B, both protocols are searched over the full nominal-duration interval.

We search over $\tau$ with mixed relative and absolute meshes, retaining disconnected success windows and refining near candidate minima and classification changes.
The new searches use a spacing no larger than $1\%$ or $5/\omega$, whichever is smaller, followed by a midpoint sweep and seven local refinement rounds near successful minima and near-threshold failures.
Stopping events are located by resolving the individual threshold crossings and testing their simultaneous satisfaction.
At point B, both protocols are also searched at nine higher joint thresholds from $0.955$ to $0.995$.
The reported $T$ is the shortest validated time found within each waveform family, with finite speed-search and spatial resolution.

The LA candidate set includes zero-floor schedules and schedules constructed with $\lambda\omega^2\mathcal A_0=20$.
At point B and 16 selected map points, we also test floors of $0.25$, $0.5$, $1$, $2$, and $4$ times the coupling-averaged indicator over $0\leq g\leq2\omega$, together with the constant-weight limit $\mathcal A_0\geq\max_g\mathcal A(g)$.
This limit gives a nearly constant-speed ramp with the same endpoint smoothing.
The B search additionally includes factors $0.125$, $8$, and $16$.
At the remaining 294 points where Linear is faster in the initial comparison, we test the constant-weight limit and refine the allocation floor whenever it improves the stopping time.
At nine points with no LA success, we test 64 logarithmically spaced durations across the nominal search interval for each of three floors: $0.5$ and $1$ times the averaged indicator and the constant-weight limit.
Any new success is then subject to the full duration search and local floor refinement.
Validated earlier waveforms remain in the candidate set throughout these refinements.

The map contains 5535 points: 3089 grid nodes of spacing $0.01$ and 2446 supplementary points in $\Omega/\omega\in[0,1]$ and $\epsilon/\omega\in[0,0.5]$.
Static coupling requirements exclude 87 points; both protocols are evaluated at the remaining 5448.
Refinement targets equal-time and success-category boundaries and preparation-time windows, with selected spacings of $0.005$ and $0.0025$.
The zero-bias row and equality in Eq.~\eqref{eq:criterion} are excluded.
At $\epsilon/\omega=0.5$, the second triplet eigenstate at finite coupling continues the interior target branch; its large-coupling endpoint is degenerate, as discussed in Sec.~\ref{sec:static}.
Each point has its own spectrum and static qualification, and each dynamically eligible point has its own LA indicator and propagation.

Waveform construction uses 1801 spectral and 2401 waveform nodes for the coarse map, and 7201 and 9601 for added samples.
The coarse-map grid checks are supplemented by checks at 191 global samples and all successful LA waveforms at 119 boundary points, using up to 28801 spectral and 38401 waveform nodes.
All successful LA waveforms at 2304 of the supplementary points are checked with 7201/9601, 14401/19201, and 28801/38401 spectral/waveform nodes.
Grid-sensitive stopping events are re-searched and verified on finer grids.

For the additional LA shape searches and the higher-threshold calculations at B, we use 28801 spectral and 38401 waveform nodes, with checks at 57601/76801 and 115201/153601 nodes.
At each improved point, two successive refinements of the allocation floor change the shortest time by less than $0.05\%$, and a fourfold refinement of the duration mesh also changes it by less than $0.05\%$.

\subsection{Truncation and propagation checks}

We solve the time-dependent Schrödinger equation in the full oscillator--triplet space of dimension $3n_{\rm ph}$ using fourth-order commutator-free propagation and Chebyshev evaluation of each exponential.
The initial map searches use 32 Fock states, with cutoff-sensitive protocols re-searched at 64 states; the added LA shape searches start with 64 states.
Stopping events are checked with 32, 64, and 96 states as the maximum step $\omega\Delta t$ decreases from $0.0625$ to $0.03125$ and $0.015625$; sensitive cases use 64, 96, and 128 states.

Every reported successful stopping state is checked by independent propagation with at least 64 Fock states; the added LA waveforms are checked with 96 states.
Figure~\ref{fig:trajectory_diagnostics} uses full 96-state trajectories at point B.
Stopping events are checked through their residuals, local slopes, and crossing-time uncertainty; full-path sampling is refined where crossing counts disagree.
Table~\ref{tab:fidelity_convergence} reports the four representative LA waveforms.

\begin{table}[tbp]
\caption{Checks of the LA waveforms at the joint threshold $0.95$.
$\omega\delta T$ compares stopping events with 64 and 96 retained Fock states and maximum steps $0.03125$ and $0.015625$.
$\delta F$ is the largest endpoint difference among the three preparation fidelities between the accepted propagator and independent DOP853 evolution with 64 or 96 states.}
\label{tab:fidelity_convergence}
\begin{ruledtabular}
\begin{tabular}{lcc}
$(\Omega/\omega,\epsilon/\omega)$ & $\omega\delta T$ & $\delta F$ \\
\hline
$(0.65,0.30)$ & $1.4\times10^{-6}$ & $2.1\times10^{-9}$ \\
$(0.35,0.40)$ & $1.7\times10^{-8}$ & $8.4\times10^{-11}$ \\
$(1,0.025)$ & $6.0\times10^{-8}$ & $2.3\times10^{-11}$ \\
$(0.125,0.49)$ & $1.6\times10^{-6}$ & $3.2\times10^{-9}$ \\
\end{tabular}
\end{ruledtabular}
\end{table}

For the independent static classification in Fig.~\ref{fig:criterion_map}(c), branch continuation starts from the exact state with largest overlap with the large-$g$ $D_0$ reference.
The coupling step is reduced when adjacent-state overlap is insufficient or an ordered level is skipped.
Checks at 643 boundary and control points with 32 and 48 states and two maximum step sizes give identical classifications.
Three sensitive assignments are further confirmed with 64 states and an anchor shifted from $2.5\omega$ to $3\omega$.
The numerical classification agrees with the analytic boundary at all displayed points.

\subsection{Comparison with reported preparation protocols}
\label{app:protocol_comparison}

We independently propagate selected schedules with their original initial states, durations, and controls.
All quoted fidelities trace out the oscillator and use the respective Bell targets, with no added target-conversion pulses.
We use the nonlinear and Stark-assisted paths in Figs.~2 and 5 of Ref.~\cite{Xie2023UltrafastBell}, starting from the oscillator vacuum with both spins up.
We use the coupling and qubit-frequency paths at their reported durations, without optimizing them again.
The nonlinear path has an unbounded initial slope; the Stark-assisted path has $\dot g_c/\omega^2\simeq0.1443$ and fixed Stark coefficients $U_1/\omega=2/3$, $U_2/\omega=1/3$.
Here $g_c$ denotes the counterrotating coupling in that model.
The nonlinear and Stark-assisted schedules give fidelities of $0.9909$ at $\omega T=34$ and $0.9925$ at $\omega T=9.8$, respectively.

For Ref.~\cite{Petiziol2019}, we use the tangent sweep with effective counterdiabatic driving, amplitude parameter $k=3$, a 7-GHz carrier, and base coupling $g/(2\pi)=50$ MHz.
Its initial dressed eigenstate is dominated by a one-photon spin product state.
This schedule gives a Bell fidelity of $0.9961$ in 130 ns, corresponding to $\omega_r T\simeq6698$ with $\omega_r$ denoting its resonator frequency.
For context, the same initial state under the unassisted polynomial detuning sweep reaches a reduced Bell fidelity of $0.9967$ at $2500$ ns.

The present route keeps both qubit parameters fixed and varies only the collective coupling, whereas these accelerated schemes coordinate several Hamiltonian coefficients.
Single-parameter control also occurs in the unassisted detuning sweep of Ref.~\cite{Petiziol2019}.

For the nonlinear path, 24 and 48 Fock states give Bell fidelities agreeing within $10^{-12}$; the Stark path is checked up to 128 states.
The selected Jaynes--Cummings Hamiltonian of Ref.~\cite{Petiziol2019} preserves a four-dimensional two-excitation subspace.
Refining its time integration and control interpolation changes the accelerated-example Bell fidelity by less than $4.2\times10^{-11}$.

\begin{figure}[t]
\centering
\includegraphics[width=3.38in]{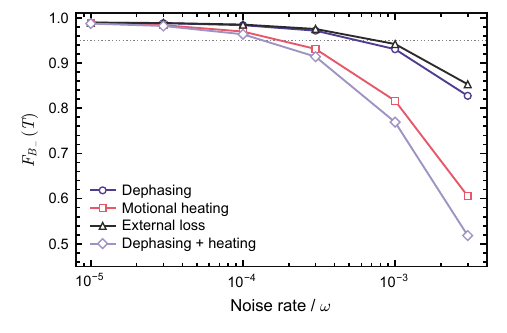}
\caption{Noise sensitivity of the fixed LA waveform at point B with coherent joint threshold $0.99$, $\omega T=96.66$, and $g(T)/\omega=1.596$.
The ordinate is the reduced-spin Bell fidelity at the prescribed stop, including loss probability.
The horizontal variable is $\gamma_\phi/\omega$ for dephasing (circles), $\kappa/\omega$ for motional diffusion (squares), and $\Gamma/\omega$ for external loss (triangles).
Diamonds show simultaneous dephasing and diffusion with equal rates and $\Gamma=0$.
The dotted guide marks $0.95$; the zero-rate value is $0.9900$.
}
\label{fig:noise}
\end{figure}

\section{Noise sensitivity of the LA preparation waveform}
\label{app:noise}

We use the LA waveform at point B that reaches the coherent joint threshold $0.99$ in Fig.~\ref{fig:trajectory_diagnostics}(d).
The schedule follows Eq.~\eqref{eq:la_schedule} with $\omega^2\mathcal A_0=3.38826$ and nominal duration $\omega\tau=108.5348$.
It stops at $\omega T=96.6643$, where $g(T)/\omega=1.59586$.
The initial state, waveform, and stopping time are held fixed as the noise rates are varied.

For the trapped-Rydberg-ion encoding of Ref.~\cite{Hamlyn2024}, the Rydberg population operator becomes $N_{rj}=(\mathbb I+\sigma_j^x)/2$ in the rotated spin convention of Eq.~\eqref{eq:Hamiltonian}.
We consider independent dephasing of this population, motional diffusion, and loss from the Rydberg level to states outside the encoded qubit space.
The density matrix restricted to the encoding obeys
\begin{align}
\dot\rho={}&-i[H(t),\rho]
+\frac{\gamma_\phi}{2}\sum_{j=1}^{2}\mathcal D[\sigma_j^x]\rho\notag\\
&+\kappa\bigl(\mathcal D[a]+\mathcal D[a^\dagger]\bigr)\rho
-\frac{\Gamma}{2}\sum_{j=1}^{2}\{N_{rj},\rho\},
\label{eq:noise_model}
\end{align}
where $\mathcal D[L]\rho=L\rho L^\dagger-\{L^\dagger L,\rho\}/2$.
Here $T_\phi=1/\gamma_\phi$ is the single-ion dephasing time in the Rydberg-population basis.
In the absence of loss, the diffusion term gives the noise contribution $d\langle a^\dagger a\rangle/dt=\kappa$ for an untruncated oscillator; $\kappa$ is therefore a heating rate in quanta per unit time.
The last term is the encoding block of a trace-preserving loss model with external sink states and no return; $1/\Gamma$ is the lifetime against that loss channel.
We retain the full two-spin space because independent dephasing can populate the singlet.
When loss is present, the trace of this block is the survival probability.
The Bell fidelity and branch fidelity are evaluated without renormalizing $\rho$, with loss events counted as unsuccessful preparation.

Figure~\ref{fig:noise} shows the final Bell fidelity for six rates from $10^{-5}\omega$ to $3\times10^{-3}\omega$.
At equal rate, the motional diffusion channel produces the largest reduction among the individual channels.
Simultaneous dephasing and heating reduce the fidelity further.
At $\gamma_\phi=\kappa=\Gamma=10^{-4}\omega$, a separate calculation including all three channels yields $F_{B_-}=0.9588$, $F_{\rm br}=0.9586$, and survival probability $0.9950$.
Together with the target eigenstate's Bell fidelity of $0.9900$, these values satisfy the joint $0.95$ preparation criterion.

The plotted density-matrix calculations use 48 oscillator states.
Repeating each channel at rates $3\times10^{-4}\omega$ and $3\times10^{-3}\omega$ with 64 states changes the final Bell fidelity, branch fidelity, singlet population, or trace by less than $1.6\times10^{-10}$; the three-channel test above satisfies the same bound.
A separately assembled sparse Liouvillian propagated with an independent integration method agrees within $3.5\times10^{-11}$ at $3\times10^{-4}\omega$ for each individual and dephasing--heating case, and at $10^{-4}\omega$ for the three-channel case.
The largest monitored population in the highest four oscillator states of an accepted calculation is below $4.7\times10^{-9}$.
The zero-rate master equation reproduces the pure-state propagation with $F_{B_-}=0.9900$ and $F_{\rm br}=0.99326$.

%

\end{document}